  \documentclass[structabstract]{aa}

\usepackage{graphicx}
 \usepackage[]{natbib}
\usepackage{txfonts}
\usepackage{color}
\definecolor{ladrillo}{RGB}{102, 0.0, 0.}
\usepackage{booktabs}
\usepackage{url}
\bibpunct{(}{)}{;}{a}{}{,} % to follow the A&A style
\begin{document}

\title{The Gaia-ESO Survey: Separating disk chemical substructures with cluster models\thanks{Based on data products from observations made with ESO Telescopes at the La Silla Paranal Observatory under programme ID 188.B-3002. These data products have been processed by the Cambridge Astronomy Survey Unit (CASU) at the Institute of Astronomy, University of Cambridge, and by the FLAMES/UVES reduction team at INAF/Osservatorio Astrofisico di Arcetri. These data have been obtained from the Gaia-ESO Survey Data Archive, prepared and hosted by the Wide Field Astronomy Unit, Institute for Astronomy, University of Edinburgh, which is funded by the UK Science and Technology Facilities Council.}}

    \subtitle{Evidence of a separate evolution in the metal-poor thin disk}
\author{A.~Rojas-Arriagada \inst{\ref{inst1}}
  \and A.~Recio-Blanco \inst{\ref{inst1}}
  \and P.~de Laverny \inst{\ref{inst1}}
  \and M.~Schultheis \inst{\ref{inst1}}
  \and G.~Guiglion \inst{\ref{inst1}}
  \and \v{S}.~Mikolaitis \inst{\ref{inst2}}
  \and G.~Kordopatis \inst{\ref{inst3}}
  \and V.~Hill \inst{\ref{inst1}}
  \and G.~Gilmore \inst{\ref{inst9}}
  \and S.~Randich \inst{\ref{inst10}}
  \and E.~J.~Alfaro \inst{\ref{inst8}}
  \and T.~Bensby \inst{\ref{inst7}} 
  \and S.~E.~Koposov \inst{\ref{inst9},\ref{inst11}}
  \and M.~T.~Costado \inst{\ref{inst8}}
  \and E.~Franciosini \inst{\ref{inst10}}
  \and A.~Hourihane \inst{\ref{inst9}}
  \and P.~Jofr\'e \inst{\ref{inst9}}
  \and C.~Lardo \inst{\ref{inst5}}
  \and J.~Lewis \inst{\ref{inst9}}
  \and K.~Lind \inst{\ref{inst12}}
  \and L.~Magrini \inst{\ref{inst10}}
  \and L.~Monaco \inst{\ref{inst6}}
  \and L.~Morbidelli \inst{\ref{inst10}}
  \and G.~G.~Sacco \inst{\ref{inst10}}
  \and C.~C.~Worley \inst{\ref{inst9}}
  \and S.~Zaggia \inst{\ref{inst4}}
  \and C.~Chiappini \inst{\ref{inst3}}
}
\institute{
Laboratoire Lagrange, Universit\'e Côte d'Azur, Observatoire de la C\^ote d'Azur, CNRS, Blvd de l'Observatoire, CS 34229, 06304 Nice cedex 4, France \email{arojas@oca.eu} \label{inst1}
\and
Institute of Theoretical Physics and Astronomy, Vilnius University, A. Go\v{s}tauto 12, 01108 Vilnius, Lithuania \label{inst2}
\and
Leibniz-Institut f\"{u}r Astrophysik Postdam (AIP), An der Sternwarte 16, 14482 Postdam, Germany \label{inst3}
\and
Institute of Astronomy, University of Cambridge, Madingley Road, Cambridge CB3 0HA, United Kingdom \label{inst9}
\and 
INAF - Osservatorio Astrofisico di Arcetri, Largo E. Fermi 5, 50125, Florence, Italy \label{inst10}
\and
Instituto de Astrof\'{i}sica de Andaluc\'{i}a-CSIC, Apdo. 3004, 18080 Granada, Spain \label{inst8}
\and 
Lund Observatory, Department of Astronomy and Theoretical Physics, Box 43, SE-221 00 Lund, Sweden \label{inst7}
\and
Moscow MV Lomonosov State University, Sternberg Astronomical Institute, Moscow 119992, Russia \label{inst11}
\and 
Astrophysics Research Institute, Liverpool John Moores University, 146 Brownlow Hill, Liverpool L3 5RF, United Kingdom \label{inst5}
\and
Department of Physics and Astronomy, Uppsala University, Box 516, SE-751 20 Uppsala, Sweden \label{inst12}
\and
Departamento de Ciencias Fisicas, Universidad Andres Bello, Republica 220, Santiago, Chile \label{inst6}
\and
INAF - Padova Observatory, Vicolo dell'Osservatorio 5, 35122 Padova, Italy \label{inst4}
 }

   \date{Received...; accepted...}

   \newcommand{\teff}{$T_{\rm eff}~$}
   \newcommand{\logg}{$\log{g}~$}
   \newcommand{\feh}{$\rm [Fe/H]~$}
   \newcommand{\met}{${\rm [M/H]}~$}
   \newcommand{\aabun}{${\rm [\alpha/Fe]}$}
   \newcommand{\kms}{km~s$^{-1}$}
   \newcommand{\vrad}{${\rm V_{rad}}$}

% \abstract{}{}{}{}{} 
% 5 {} token are mandatory
 
  \abstract
  % context heading (optional)
  % {} leave it empty if necessary  
   {Recent spectroscopic surveys have begun to explore the Galactic disk system on the basis of large data samples, with spatial distributions sampling regions well outside the solar neighborhood. In this way, they provide valuable information for testing spatial and temporal variations of disk structure kinematics and chemical evolution.}
  % aims heading (mandatory)
   {The main purposes of this study are to demonstrate the usefulness of a rigorous mathematical approach to separate  substructures of a stellar sample in the abundance-metallicity plane, and provide new evidence with which to characterize the nature of the metal-poor end of the thin disk sequence.}  
  % methods heading (mandatory)
   {We used a Gaussian mixture model algorithm to separate in the [Mg/Fe] vs. [Fe/H] plane a clean disk star subsample (essentially at ${\textmd R}_{\textmd GC}<10$ kpc) from the Gaia-ESO survey (GES) internal data release 2 (iDR2). We aim at decomposing it into data groups highlighting number density and/or slope variations in the abundance-metallicity plane. An independent sample of disk red clump stars from the \textit{Apache Point Observatory Galactic Evolution Experiment} (APOGEE) was used to cross-check the identified features.}
  % results heading (mandatory)
   {We find that the sample is separated into five groups associated with major Galactic components; the metal-rich end of the halo, the thick disk, and three subgroups for the thin disk sequence. This is confirmed with the sample of red clump stars from APOGEE. The three thin disk groups served to explore this sequence in more detail. The two metal-intermediate and metal-rich groups of the thin disk decomposition ([Fe/H]$>-0.25$ dex) highlight a change in the slope at solar metallicity. This holds true at different radial regions of the Milky Way. The distribution of Galactocentric radial distances of the metal-poor part of the thin disk ([Fe/H]$<-0.25$ dex) is shifted to larger distances than those of the more metal-rich parts. Moreover, the metal-poor part of the thin disk  presents indications of a scale height intermediate between those of the thick and the rest of the thin disk, and it displays higher azimuthal velocities than the latter.  These stars might have formed and evolved in parallel and/or dissociated from the inside-out formation taking place in the internal thin disk. Their enhancement levels might be due to their origin from gas pre-enriched by outflows from the thick disk or the inner halo. The smooth trends of their properties (their spatial distribution with respect to the plane, in particular) with [Fe/H] and [Mg/Fe] suggested by the data indicates a quiet dynamical evolution, with no relevant merger events. }
  % conclusions heading (optional), leave it empty if necessary 
   {}

   \maketitle
   
%==================================================================================================
%     Introduction
%==================================================================================================
\section{Introduction}
\label{sec:introduccion}

A significant new piece in our understanding of the global structure of the Milky Way was added around 20 years ago with the discovery of the thick disk \citep{yoshii1982,gilmore1983}. Detected from a double exponential fit to the vertical number density distribution of disk stars, the thick disk was first characterized as a distinct structural component, with a larger scale height and hotter kinematics than the younger thin disk. Subsequent works have characterized it with kinematical and chemical properties in between those of the halo and thin disk populations \citep{feltzing2003,reddy2006,navarro2011,kordopatis2011,kordopatis2013b,kordopatis2013a,fuhrmann1998,fuhrmann2004,fuhrmann2011,bensby2014,recio-blanco2014}. However, a large part of the observational work aiming at characterizing the two disks resides in samples of stars located in the solar neighborhood. It has been possible only recently to begin to extend the spatial coverage of samples, spanning larger radial regions with the SEGUE \citep{yanny2009}, RAVE \citep{steinmetz2006}, APOGEE \citep{eisenstein2011}, and GES surveys \citep{GESMessenger}.

A main question in all these studies is the way the thin and thick disk samples are defined and separated. The adopted criterion is, at the same time, a definition of what we understand by thin and thick disk, and has an impact on the distribution of other properties characterizing them. The chosen parameter or criterion should allow describing the large-scale distribution of stars in both disks. Star-count-based characterizations of the disk system are prone to systematics arising from degeneration in age-color relations or from the adoption of a single scale height to model the thick disk in all its radial extension. These effects can lead to spurious results, blurring other properties by the cross contamination of thick with thin disk stars in the samples. A kinematical selection \citep{soubiran2003, bensby2003} allows obtaining cleaner distributions in other parameters, but some superposition still exists. Recently, it has been proposed to use chemical information to separate disk samples from the relative enhancement in $\alpha$-elements \citep[e.g.,][]{navarro2011,adibekyan2013,recio-blanco2014,mikolaitis2014,kordopatis2015b}. The chemical composition of a star is a more stable property than the kinematics or spatial distributions during the complex evolution of a stellar system \citep{freeman2002}. The idea is to define thin and thick disk stars from their location in the $\alpha_\textmd{abundance}$-metallicity plane, which is conceptually equivalent to understanding them as dual different star formation rates \citep{chiappini2009}. When the data quality and the resolution are good enough, two distinct sequences of $\alpha$-poor and $\alpha$-rich stars, associated with the thin and thick disks, allow separating samples with remarkable results in spatial, kinematics, and age distributions.

In a different interpretation of SEGUE data, \citet{bovy2012} discarded the existence of a thick disk as a different structural component. There is no necessity of a double exponential to explain the vertical stellar density profile in their mass-weighted scale-height distribution. In this sense, they interpreted the observed bimodality in the [$\alpha$/Fe] vs. [Fe/H] plane as an effect of the lack of correction of the spectroscopic sample to account for the survey selection function, otherwise affected by the survey-specific spatial and mass sampling of the underlying stellar populations.

\citet{anders2014}, using the first year of APOGEE data, have confirmed the existence of a dip/gap (a low density region in the number density distribution) in the [$\alpha$/Fe] vs. [Fe/H] plane, not only in the solar distance bin, but also in two other radial bins in the inner and outer parts of the galactic disk. These authors argued that in APOGEE the target selection would not be enough to create such a gap. This was later on confirmed by \citet{nidever2014} with a red clump (RC) sample for which the uncertainties in the distances could be reduced from the typical ~20\% uncertainties to $\sim5$\%, as claimed in \citet{bovy2014}.

In the thin and thick disk sequences framework, ages were determined by \citet{haywood2013} to further characterize the HARPS local neighborhood sample of \citet{adibekyan2013}. They identified the thin and thick disk sequences  with two epochs in the formation history of the Milky Way, with different timescales, different enrichment rates, and different assembling mechanisms. Their age-based classification
into thin and thick disk agrees remarkably well with the location of those stars in the two sequences of the abundance-metallicity plane.

Using a sample of $\sim10000$ RC stars from the SDSS-III/APOGEE, \citet{nidever2014} traced patterns in the [$\alpha$/Fe] vs. [Fe/H] plane through a large radial domain. The thick disk displays a relatively constant trend in this plane with radial distances, while the thin disk distribution is more radial dependent. They suggested that while in the inner Galaxy stars in both disks could be explained  by a single chemical evolutionary track, more complicated assumptions are needed to explain their distributions in outer regions. These results were confirmed over a larger distance range in \citet{hayden2015} by using a sample of red giant stars from APOGEE DR12.

Other recent studies have characterized the thick disk as a structure with negative vertical and flat radial metallicity gradients \citep{mikolaitis2014}. The thin disk instead presents gradients in both directions \citep{hayden2014,recio-blanco2014,mikolaitis2014}, an imprint predicted by the inside-out disk formation paradigm
that was first proposed by \citet{larson1976}.

In the same line, observations of open clusters as tracers of young stellar populations (which makes them a good proxy for the current interstellar medium (ISM) metallicity) show that outside the solar annulus, at distances R$_{\textmd GC}>9-10$ kpc and subsolar metallicity, the radial metallicity gradient flattens \citep{bragaglia2012,yong2012,frinchaboy2013}. This shallow slope is consistent with determinations based on Cepheids \citep{genovali2014}. In this context, the existence of an external distinct region of the thin disk (with $\textmd{[Fe/H]<-0.25}$ dex) has been proposed. From age and kinematics considerations, \citet{haywood2013} has proposed a different chemodynamical history for this region, whose properties seem to be intermediate to those of the internal thin and thick disk. The presence of such metal-poor thin disk stars in their local sample is explained in this scenario as an effect of the Sun at the interface of the two regions of the disk \citep{snaith2015}, so without the necessity of invoking churning, but maybe some degree of blurring radial migration processes.

All in all, the interpretation of the stellar distribution in the metallicity-abundance plane in the context of disk structures is a current hot topic of debate. Separating the relative contribution of thick/thin, internal/outer disks, and the variations in their properties with R$_{\textmd GC}$ and Z is an important step to reconstruct the formation scenario of the disk as a complex entity during the assembly of the Milky Way.

In this context, we here use a sample of stars\footnote{Most of them are located in the spatial region given by $5<{\textmd R}_{\textmd GC}<10$ kpc and $\left|Z\right|<2$ kpc.} selected from the iDR2 Gaia-ESO survey \citep{GESMessenger} to study the substructures of the Galactic disks from their distribution in the [Mg/Fe] vs. [Fe/H] plane. So far, almost no statistically rigorous approach has been proposed to cope with this problem \citep[but see][]{kordopatis2015b}. For this purpose, we introduce an objective criterion to separate sequences into the abundance-metallicity plane by using a Gaussian mixture model (GMM) clustering algorithm. In the next section we present the iDR2 Gaia-ESO survey and APOGEE data sample selection. The latter, having a very different selection function, is used to cross validate our cluster analysis. In Sect. \ref{sec:data_structure} we explore the data distribution in the [Mg/Fe] vs. [Fe/H] plane. The GMM decomposition of the disks and their characterization is presented in Sect. \ref{sec:decomp_ges_sample}. The metal-poor end of the thin disk sequence is further explored in Sect. \ref{sec:metal-poor_thin_disk}. Finally, we summarize and discuss our main results and conclusions in Sect. \ref{sec:discusion} and \ref{sec:conclusiones}.

%==================================================================================================
%     Data samples
%==================================================================================================
\section{Data samples}
\label{sec:datos}
For the present work we made use of  data coming from  the Gaia-ESO \citep{GESMessenger} and APOGEE \citep{eisenstein2011} spectroscopic surveys. In both cases we only considered lines of sight corresponding to disk fields, excluding clusters and  pointings toward the bulge. 
GES data were acquired from the second internal Data Release (iDR2). It comprises a subset of 8906 FGK field stars observed with the VLT/GIRAFFE spectrograph with both the HR10 (5339-5619\AA, R$\sim19800$)  and HR21 (8484-9001\AA, R$\sim16200$) setups. Radial velocities were computed using a spectral fitting technique through cross correlation against real and synthetic spectra (see Koposov et al 2015, in preparation). Typical  uncertainties are of around of 0.3-0.4 \kms.

Fundamental parameters T$_{\textmd{eff}}$, $\log(g),$ and [M/H], $\alpha,$ and iron-peak abundances are determined  and compiled  by the Gaia-ESO survey working group 10 (WG10), which is in charge of the GIRAFFE spectrum analysis of F, G, and K type stars. They constitute the recommended parameters by the GES consortium. A detailed description of the process can be found in Recio-Blanco et al. 2015, in preparation. Briefly, the individual spectra are analyzed by three independent nodes using different algorithms: Spectroscopy Made Easy \citep[SME,][]{valenti1996}, FERRE \citep{allende-prieto2006}, and MATISSE \citep{Matisse}. In this way, T$_{\textmd{eff}}$, $\log(g)$, [M/H], and [$\alpha$/Fe] are determined. A set of benchmark stars \citep{jofre2014} is also analyzed, and the dispersion and bias of the obtained parameters with respect to the nominal ones is calculated. Given that the smallest zero-point offsets with respect to the benchmarks are obtained from the SME analysis, this scale is taken as a reference. The results of each node are tranformed into this reference scale. When the results are in the SME scale, they are corrected by removing the zero points with respect to the benchmarks. Finally, the results from the different nodes are combined on average to produce a unique set of atmospheric parameters while reducing the random errors of individual determinations.

Adopting this set of final parameters, elemental abundances are determined by three independent algorithms: SME, an automated spectral synthesis method \citep{mikolaitis2014}, and a Gauss-Newton  method using a precomputed grid of synthetic spectra \citep{guiglion2014}. Several $\alpha$ and iron-peak abundances (including the iron and magnesium abundances used in this work) are determined in this way, adopting the Gaia-ESO survey line list and the MARCS model atmospheres \citep{gustafsson2008}. The results of the different nodes compare well, and only small shifts are necessary to place them on the same scale. These shifts are independent of atmospheric parameters, so that the final abundances for each element are calculated for each star as the average of the individual determinations. The final abundances relative to the Sun are derived by adopting the solar composition of \citet{grevesse2007}.

Distances were obtained via isochrone fitting using the method described in \citet{kordopatis2011} and \citet{schultheis2015}.

\subsection{Adopted GES sample}
\label{subsec:ges_final_sample}

As explained in \citet{recio-blanco2014}, the GES target selection for halo and disk is based on the disk-to-halo transition seen in SSDS at $17<r<18$ and $0.2<g-r<0.4$. Guided by this feature, a selection function based on VISTA photometry is defined in two nominal boxes:

\begin{itemize}
 \item Blue box: $0.0<\textmd{J-K}<0.45$; $14.0<\textmd{J}<17.5$
 \item Red box: $0.4<\textmd{J-K}<0.70$; $12.5<\textmd{J}<15.0.$
\end{itemize}

Their effective positions are adjusted according to the field extinction, as given by the Schlegel extinction maps \citep{schlegelMapas}.

\begin{figure*}
\begin{center}
\includegraphics[width=12cm]{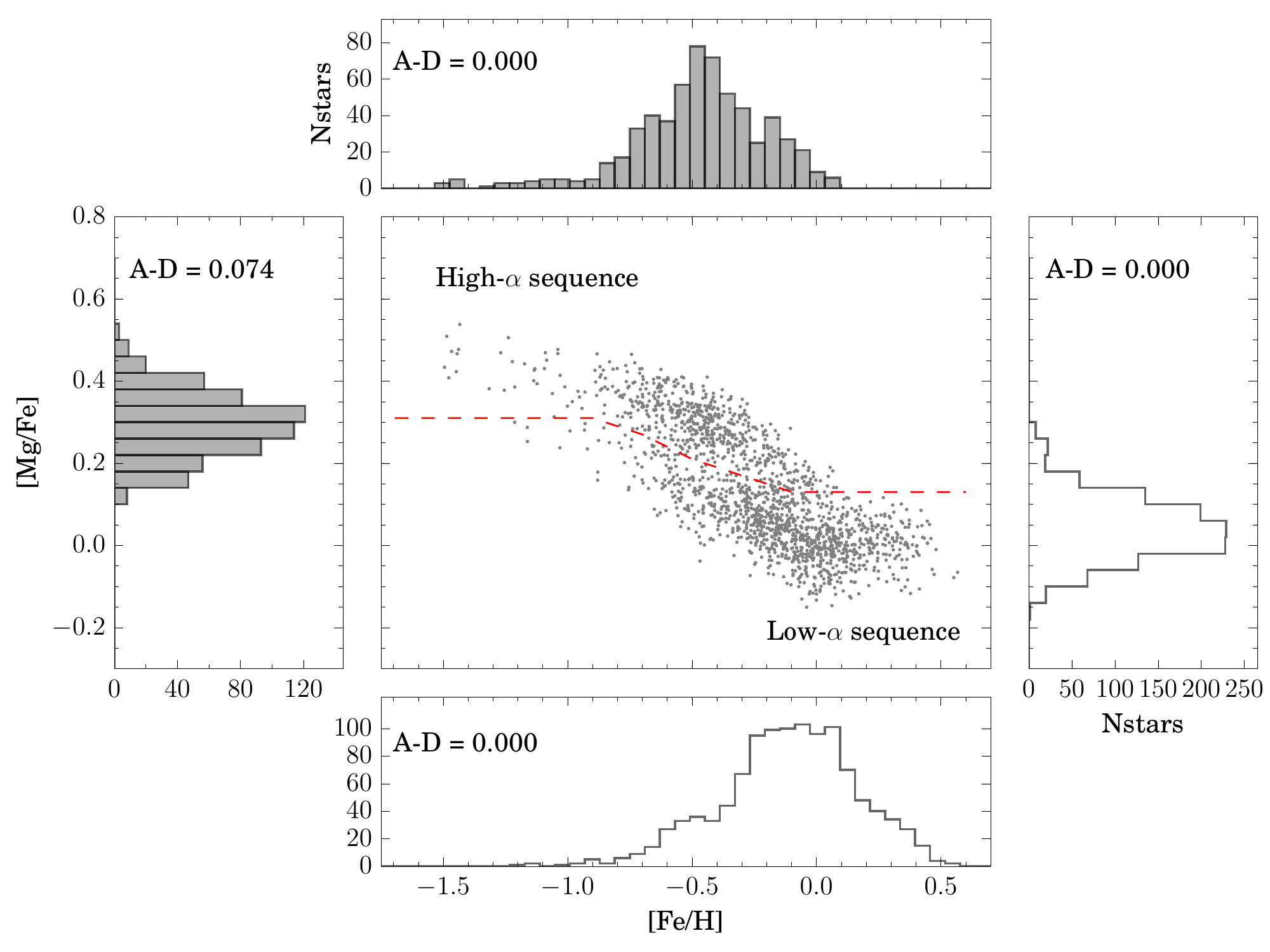}
\caption{Final clean GES dataset of 1725 stars in the [Mg/Fe] vs. [Fe/H] plane. A red dashed line marks the separation between $\alpha$-rich and $\alpha$-poor (thick and thin disks) estimated as the minimum of the [Mg/Fe] distributions in several intervals of metallicity (as made in \citeauthor{recio-blanco2014} \citeyear{recio-blanco2014} and  \citeauthor{mikolaitis2014} \citeyear{mikolaitis2014} in the [$\alpha$/Fe] vs. [M/H] plane). The four panels around the central panel display the marginalized distributions in metallicity and Mg abundance corresponding to these two disk sequences. Bar histograms stand for the thick disk and step histograms for the thin disk. A $p-$value, derived from the Anderson-Darling test for Gaussianity, is quoted in each panel.} 
\label{fig:data_overview}
\end{center}
\end{figure*}

The GES dataset contains stars in 129 different lines of sight at high- and low-latitude fields. From the characteristics of the selection function, we expect the data to be composed of disk(s) and halo stars. Only the best-quality stellar parametrizations were kept by considering the smallest errors in fundamental parameters and elemental abundances. These errors (the node-to-node dispersion of the atmospheric parameters) come from the homogenization process between GES nodes used to calculate the final set of recommended parameters, as described above, and provide an idea of the stability of the results for a given star. As another selection criterion, we  used the signal-to-noise ratio (S/N)of the originally analyzed spectrum in the HR10 setup of GIRAFFE. The corresponding S/N in the HR21 setup is about twice as high. To clean up the sample, we applied cuts to the 90$^{th}$ percentile of the error distributions in stellar parameters, magnesium abundance, and metallicity dispersions. These cuts correspond to $\Delta$\teff=136 K, $\Delta$\logg=0.27 dex, $\Delta$[\ion{Fe}{I}/H]=0.14 dex, and $\Delta$[\ion{Mg}{I}/H]=0.10 dex. Therefore, stars with errors simultaneously smaller than the respective cut values that have a S/N > 33 were kept. For the present analysis, we adopted a metallicity [Fe/H] and magnesium abundance ratio [Mg/Fe] as determined from \ion{Fe}{I} and \ion{Mg}{I} lines. Typical mean errors in these quantities are $\overline{\textmd{E[Fe/H]}}=0.06$ dex and $\overline{\textmd{E[Mg/Fe]}}=0.07$ dex for our selected sample. The sample presents a fairly clean distribution in the plane [Mg/Fe] vs. [Fe/H], with only a small number of outliers. To facilitate the analysis, we removed these outliers, 12 in total, which were well distributed in the metallicity range of the data, with a MAD-clipping algorithm. We also removed a few $\alpha$-poor stars with [Fe/H]<-1.5 dex that are too metal-poor to belong to the thin disk. The resulting cleaned sample, comprising 1725 stars, is shown in Fig. \ref{fig:data_overview}.

\subsection{Adopted APOGEE sample}
\label{subsec:apogee_final_sample}
APOGEE is a near-infrared (H-band) high-resolution ($R\sim22500$) spectroscopic survey carried out during bright time at the 2.5-meter Sloan telescope (Apache Point Observatory).

The APOGEE selection function is described and characterized in detail in \citet{zasowski2013}. Broadly speaking, it selects stars satisfying the color-magnitude constraints $\textmd{(J-K)}_0\geq 0.5$ and $7.0\geq \textmd{H} \geq 13.8$ by using associated 2MASS photometry. Colors are dereddened with extinction derived using the Rayleigh Jeans color excess (RJCE) method \citep{majewski2011}. The color cut was set red enough to remove main-sequence stars, but blue enough to reduce the bias against metal-poor giants.

We used the APOGEE red clump catalog (APOGEE-RC) of 19937 stars described in \citet{bovy2014}. In this work,  RC stars were selected from the general APOGEE DR12 catalog based on their position in the color-metallicity-\teff-\logg space. The classification method is calibrated using stellar evolution models and high-quality asteroseismology data. In this way, the derived distances for the sample are claimed to be accurate to 5-10 \%. 

In order to work with a clean sample, we made use of several flags provided by the apogee pipeline ASPCAP \citep{garcia-perez2014}. They characterize the quality of the spectra and fitting procedure, allowing us to remove stars with less robust parameters (in particular, we used ASPCAP $\chi^2<10$ and ASPCAPFLAGS containing a warning about any parameter as criteria to eliminate stars from the sample). We also selected stars with S/N>220. By applying all these cuts, we obtained our \textit{\textup{working sample}}, composed of 6107 RC stars with good quality stellar parameters. The ASPCAP pipeline provides \teff, \logg, [M/H], [C/M], [N/M], and [$\alpha$/M]. A calibrated version of them is obtained by empirical comparisons to external spectroscopic references \citep{meszaros2013} and is provided in the public APOGEE-RC catalog. We adopted these calibrated stellar parameters for our analysis.
We use this \textit{\textup{working sample}}, in particular [M/H] and [$\alpha$/M], in Sect. \ref{subsec:gmm_comparacion_apogee} to perform a qualitative comparison with the results obtained from Gaia-ESO survey data. Typical mean errors in these quantities are $\overline{\textmd{E[M/H]}}=0.03$ dex and $\overline{\textmd{E[}\alpha/\textmd{M]}}=0.01$ dex.

%==================================================================================================
%  Metallicity abundance distributions and analysis strategy
%==================================================================================================
\section{Metallicity-abundance distributions and analysis strategy}
\label{sec:data_structure}
We used the cleaned dataset to explore the possible chemical substructure in the plane [Mg/Fe] vs. [Fe/H]. We adopted this plane because the abundance determinations for the $\alpha$-element Mg  seem to be less affected by errors in stellar parameters and because visual inspection showed some evidence for a better
separation of  thin and thick disk than for other elements \citep[see Fig. \ref{fig:data_overview} and][]{mikolaitis2014}. Moreover, as pointed out by \citet{gonzalez2011}, together with oxygen, magnesium it is expected to be produced exclusively by SNII explosions.

Using the procedure presented in \citet{recio-blanco2014}, we separated the sample into thin and thick disk sequences. In Fig. \ref{fig:data_overview} this separation is depicted by the red dashed line. For each sequence, we constructed histograms, both in [Fe/H] and [Mg/Fe] to verify the data structure. They are displayed in the lateral panels of Fig. \ref{fig:data_overview}. A visual inspection shows that in each case the data cannot be represented by a single Gaussian distribution. In all cases, it is possible to find visual evidence for substructures, secondary peaks, and departure from Gaussianity. To quantify these departures, we performed an Anderson-Darling test in each distribution. This statistic tests for normal distribution with unknown mean and variance. The corresponding $p$-values\footnote{A $p$-value is defined as the probability of obtaining a test statistic result at least as extreme or as close to the one observed assuming that the null hypothesis $H_0$ is true. We reject $H_0$ when the $p$-value is lower than a predetermined significance level, often 0.05. Then, the observed result is highly unlikely under $H_0$.} are quoted in each panel. We clearly see that in almost every case we can reject the null hypothesis that the sample comes from a normal distribution. The only exception might be the [Mg/Fe] distribution of the thick disk, with a $p$-value=0.07, although it is fairly close to the 5\% limit.

This preliminary data overview is important to guide the adopted strategy for data analysis. As the statistical test shows, the distributions cannot be explained by a single Gaussian, but their general shape suggests a possible mixture of two or three peaked distributions. In the following we assume multiple Gaussianity as a reasonable simple approximation and verify a posteriori whether this adopted assumption satisfactorily reproduces the shape of the data, providing also physically interpretable results.

In view of the data properties revealed by the above exploratory inspection, we adopted the formalism of GMM to perform a clustering analysis searching for subgroups in the data. This unsupervised analysis approach does not require priors. A GMM is a parametric probability density function represented as a weighted sum of Gaussian component densities. The GMM parameters (including the number of components and centroids) are estimated to better represent the analyzed data structure.
Given a dataset with a certain number of features (points), a GMM algorithm thus constructs a generative model that is the specific Gaussian mixture that better predicts the data structure. Given this statistically optimal model with a specific number of clusters, we can separate the data into families according to the component that has the highest probability to have generated each feature. The cluster parameters can then be used to describe the data in terms of dispersion, correlations, or slopes.
In the following, we describe some of these points in more detail.

%-------------------------------------------------------
% \subsection{The GMM formalism}
% \label{subsec:gmm_formalismo}
A mixture model ($M$) is a weighted sum of a number ($k$) of probability distribution functions ($pdf$). For a GMM in 2D, the mixture is defined by a sum of bivariate  normal distributions ($G$)

$$M(z|\mu,\Sigma)=\sum_{i=1}^{k} w_i  G(z|\mu_i,\Sigma_i),$$

with means $\mu$ and covariance matrices $\Sigma$. The mixture model with a specific set of parameters $\mu_i$ $\sigma_i$, and weights $w_i$ for the components is an attempt to fit the data set $z$ composed by N features. Given such data set, we wish to determine the parameters of the mixture of $k$ Gaussian modes that better predict the data structure. A mixture model must satisfy

$$\sum_i^k w_i=1$$

and should integrate to 1 over the data coordinate space.

The likelihood function of a mixture is defined as
$$L(z|\mu,\Sigma)=\prod_{j=1}^{\textmd{N}}M(z_j|\mu,\Sigma).$$
Then, the log-likelihood corresponds to
\begin{align*}
\ln(L(z|\mu,\Sigma)) & = \ln\left(\prod_{j=1}^{N}\sum_{i=1}^{k} w_i  G(z_j|\mu_i,\Sigma_i) \right)\\
                     & = \sum_{j=1}^{N}\ln\sum_{i=1}^{k} w_i  G(z_j|\mu_i,\Sigma_i).
\end{align*}

The Expectation-Maximization algorithm maximizes the likelihood function (equivalent to minimize the negative of the log-likelihood), determining the set of parameters that defines the best mixture model $\hat{M}(z|\mu,\Sigma)$.

By fitting a GMM to a data set, we can characterize the individual modes of the probability distribution with their means and covariances. The Gaussian modes are a good way of clustering the data points into similar groups. When the fit is performed, it is then possible to check to which mode each data point belongs most probably, by calculating the responsibilities

$$R(i|z_j)=\frac{w_iG(z_j|\mu,\Sigma)}{\sum_m w_m G(z_j|\mu_m,\Sigma_m)}.$$\\

This means that for each datapoint there is a set of \textit{\textup{a posteriori}} probabilities to have been generated from each component of the mixture. A  decision criterion, such as the highest of such responsibilities, can be used as a label to separate the sample into several data groups. More elaborated approaches can be devised to deal with the fact that some points with close equal probabilities to belong to several components can still be classified just in one with a simple criterion.

The Expectation-Maximization algorithm therefore allows us to determine the best parameters of a mixture model with a given number of modes. In a general case problem, we normally do not know a priori the number of components into which the data substructure separates. We thus need an extra loop of optimization to compare several maximum likelihood models with different number of components.

To do this, we adopted the Akaike information criterion (AIC)\footnote{Another alternative could be the Bayesian information criterion (BIC). It  penalizes a  solution with increasing level of complexity (i.e., larger number of modes in the mixture) more strongly than the AIC.} as a cost function to assess the relative fitting quality in between different proposed mixtures. The AIC is a measure of the relative quality of a statistical model for a given set of data. It is designed to deal with the trade-off of the goodness of fit and the complexity of the model (number of GMM components). We can use it as a way to perform model selection, allowing us to determine the number of components that are presumably supported by the data. The AIC is given by
$$AIC=2N_p-2\ln(L_{max}),$$
where $N_p$ is the number of parameters of the model and $L_{max}$ is the maximized value of the likelihood function of the model (obtained through the Expectation-Maximization algorithm).

The AIC is a relative quantity in the sense that it is not a test for the null hypothesis. Indeed, it cannot tell us anything about the quality of the model in an absolute sense. The only significant information is the relative comparison of the AIC between different proposed models. As a consequence, the model with the lowest AIC value is the preferred one for a given data set. The quantity

$$\exp\left[\frac{\textmd{AIC}_{min}-\textmd{AIC}_i}{2}\right]$$

gives us the relative probability that the $i$-th model minimizes the information loss. This is the relative likelihood of the model $i$.

We performed several tests using simulated distributions trying to reproduce the possible disk sequence distributions in the abundance-metallicity plane. Under the assumption of Gaussian components made by the GMM approach, a deviation from Gaussianity, like a tail in a given distribution, will be  separated into an extra component.  If the dispersion and superposition of data substructures is not large, the method is quite efficient in recovering the characteristics of the generative model of the simulated distributions.

%==================================================================================================
%  GMM decomposition of the samples
%==================================================================================================
\section{GMM decomposition of the GES and APOGEE samples}
\label{sec:decomp_ges_sample}

We used the GMM analysis approach to scrutinize GES and APOGEE samples. We aimed at detecting and characterizing the number density and slope variations of the different sequences in the abundance-metallicity plane.

\subsection{GES data sample}
\label{subsec:gmm_ges_data_sample}
\begin{figure}
\begin{center}
\includegraphics[width=8.9cm]{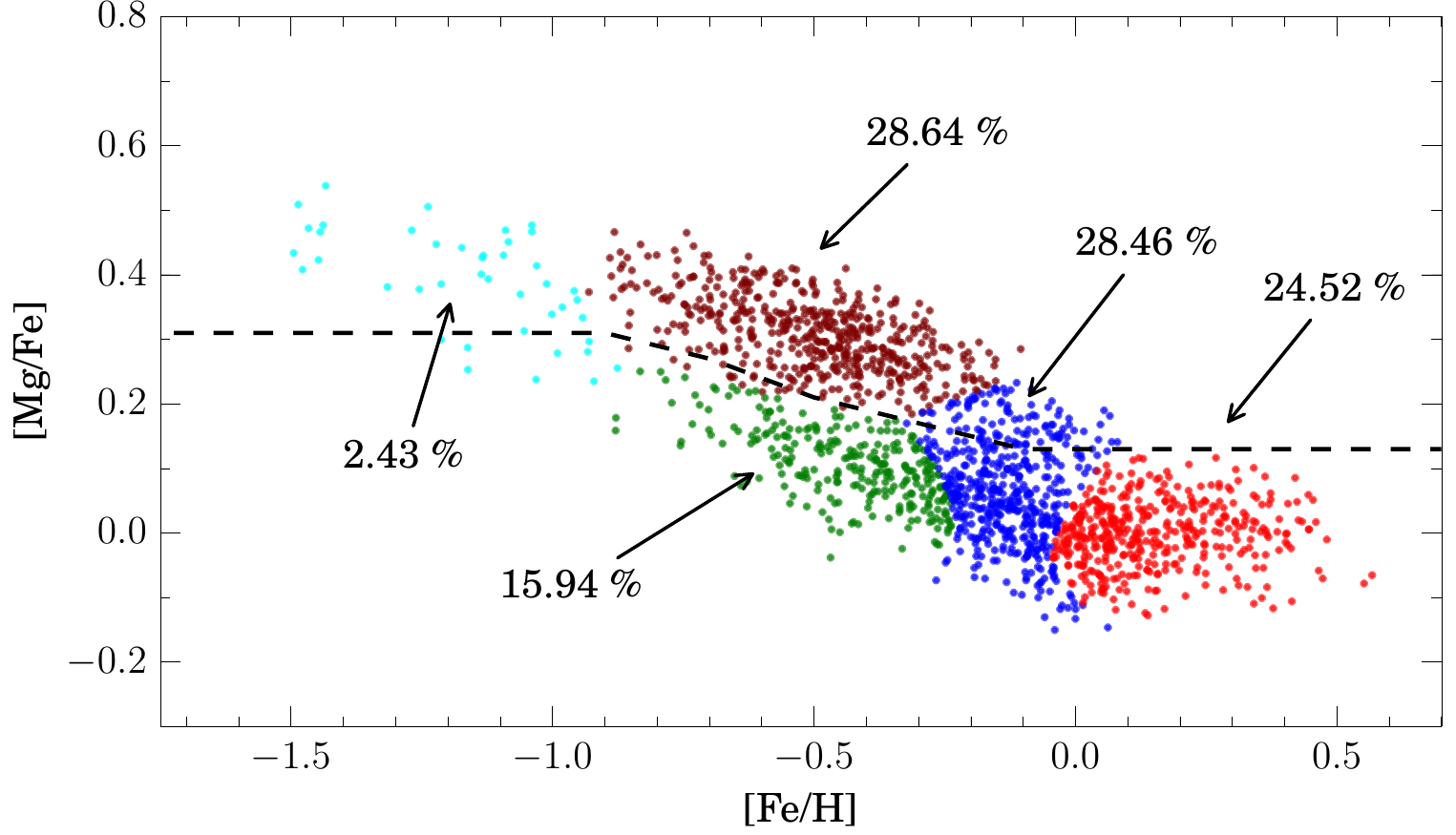}
\caption{GMM best model for the GES data distribution in the [Mg/Fe] vs. [Fe/H] plane. The colors highlight the different data families corresponding to individual modes of the Gaussian mixture. Stars are classified according to the highest component-associated responsibilities. Three black dashed line segments displays the same division into thin and thick disk sequences as in Fig. \ref{fig:data_overview}. We quote the percentage of the sample that each component encompasses.} 
\label{fig:gmm_vs_split_manual}
\end{center}
\end{figure}

We applied the GMM algorithm to our data in the [Mg/Fe] vs. [Fe/H] plane, searching for mixtures composed of up to 10 components\footnote{This is an overestimated number, since we expect the disks and halo to be represented by fewer components. This is just to ensure that the selected model based on the AIC is the preferred one considering also potentially overfitted solutions.}. To avoid overfitting, we set a minimum allowed value for the covariance of the Gaussian modes of the mixture. We adopted $\sigma_{min}=0.05$ dex, which is permissive, but avoids unphysically narrow components in the data. We wished to avoid finding substructures with a size smaller than the typical errors in [Mg/Fe] or [Fe/H] ($\overline{\textmd{E[Fe/H]}}\simeq0.06$ dex and  $\overline{\textmd{E[Mg/Fe]}}\simeq0.07$ dex). In Fig. \ref{fig:gmm_vs_split_manual} we display the best model according to the AIC model selection criterion, and for comparison, the same thin and thick disk division displayed in Fig. \ref{fig:data_overview}. The dataset is separated into five components\footnote{This number seems robust since the relative probabilities associated to the GMM solutions with 4 and 6 components are $10^{-11}$ and 0.005, respectively.}, as shown by the different colors used to visually tag the stars belonging to the modes of the GMM model (hereafter, we keep these colors in the other figures to better identify the different data groups). We recall here that this color classification is only indicative. In fact, as explained in Sect. \ref{sec:data_structure}, we classified stars into the different modes of the GMM based on the highest associated responsibility, regardless of the relative importance of other components to explain that observation. This means that stars located in regions close to the borders between components can have nearly equal probabilities to belong to two or even three of them, which makes their classification uncertain. We consider these stars in Sect. \ref{sec:metal-poor_thin_disk}. For our current qualitative description of the  GMM results, a highest probability tagging is enough. The centroids and marginalized dispersions of the Gaussian modes of the best model are listed at the left side of Table \ref{tab:gmm_results}.

Three of the GMM components (red, blue and green symbols) constitute a low-$\alpha$ thin disk sequence. A high-$\alpha$ component (brown symbols) stands for the thick disk, and a high-$\alpha$ metal-poor component (cyan symbols)  for the metal-poor tail of the thick disk and/or halo stars. The multimodal nature of the thin disk sequence qualitatively agrees with the non-Gaussianity of its marginalized distributions in [Fe/H] and [Mg/Fe] (Fig. \ref{fig:data_overview}).

Similarly, an indication of a unique strong  high-$\alpha$ component was given by the Gaussianity of the [Mg/Fe] distribution of the thick disk sequence, as defined in Fig \ref{fig:data_overview}.

The low-$\alpha$ metal-rich end of the halo GMM group merges at the metal-poor end of both disks, configuring a region where the GMM classification of features is more difficult to assess. The status of this region of the [Mg/Fe] vs. [Fe/H] plane is discussed in Sect. \ref{sec:discusion}.

Between $\textmd{[Fe/H]=-0.8}$ and $-0.3$ dex, the agreement between the two methods (GMM vs. dashed line depicting  [Mg/Fe] number density minima in metallicity bins) in separating the thin and thick disk is quite satisfactory. In contrast, the blue GMM group at $\textmd{[Fe/H]}>-0.25$ dex includes both the thin disk and what is called high $\alpha$ metal-rich stars in \citet{adibekyan2013} and \citet{gazzano2013}. It is unclear from visual inspection whether these high $\alpha$ metal-rich stars are a metal-rich extension of the thick disk or a different population. The GMM algorithm does not find a further separation accounting for such a component in GES data. The gap reported by \citet{adibekyan2013} at [Fe/H]$=-0.2$ dex that separates the thick disk sequence and the high $\alpha$ metal-rich stars is not visible in our data or in the APOGEE-RC sample (Sect. \ref{subsec:gmm_comparacion_apogee}).

%==================================================================================================
\subsection{Comparison with APOGEE-RC data}
\label{subsec:gmm_comparacion_apogee}

Our GMM analysis, illustrated in Fig. \ref{fig:gmm_vs_split_manual}, allows us to decompose the data sample into a number of subgroups
that highlight the number density and slope variations in the abundance-metallicity plane. In particular, the thin and thick disk sequences are separated by the underdensity at [Mg/Fe]$\sim0.2$ dex. This abundance gap, also seen in SEGUE data, has been characterized as an artifact arising from a sampling that is biased by the
survey selection function \citep{ivezic2008,bovy2012}.
\citet{anders2014} first confirmed the existence of the gap using APOGEE first-year data, suggesting that selection effects cannot account for the dip in the data. This was later confirmed by \citet{nidever2014}, who expanded the three radial bins discussed in \citet{anders2014} to more bins in the Z direction.

Sampling bias effects are difficult to characterize for a given survey. On the other hand, a comparison between results coming from two different surveys, with different selection functions, can shed light on the relevant factors affecting the homogeneous sampling of the underlying stellar populations, and their relative impact.

With this idea in mind, we compared the results presented in the previous subsection with an equivalent analysis carried out on the APOGEE-RC \textit{\textup{working sample}} presented in Sect. \ref{subsec:apogee_final_sample}. For this analysis, we analyzed the APOGEE sample in the [$\alpha$/M] vs. [M/H] plane (which are parameters calibrated with respect to external spectroscopic references, \citeauthor{meszaros2013} \citeyear{meszaros2013}).

\begin{figure}
\begin{center}
\includegraphics[width=8.9cm]{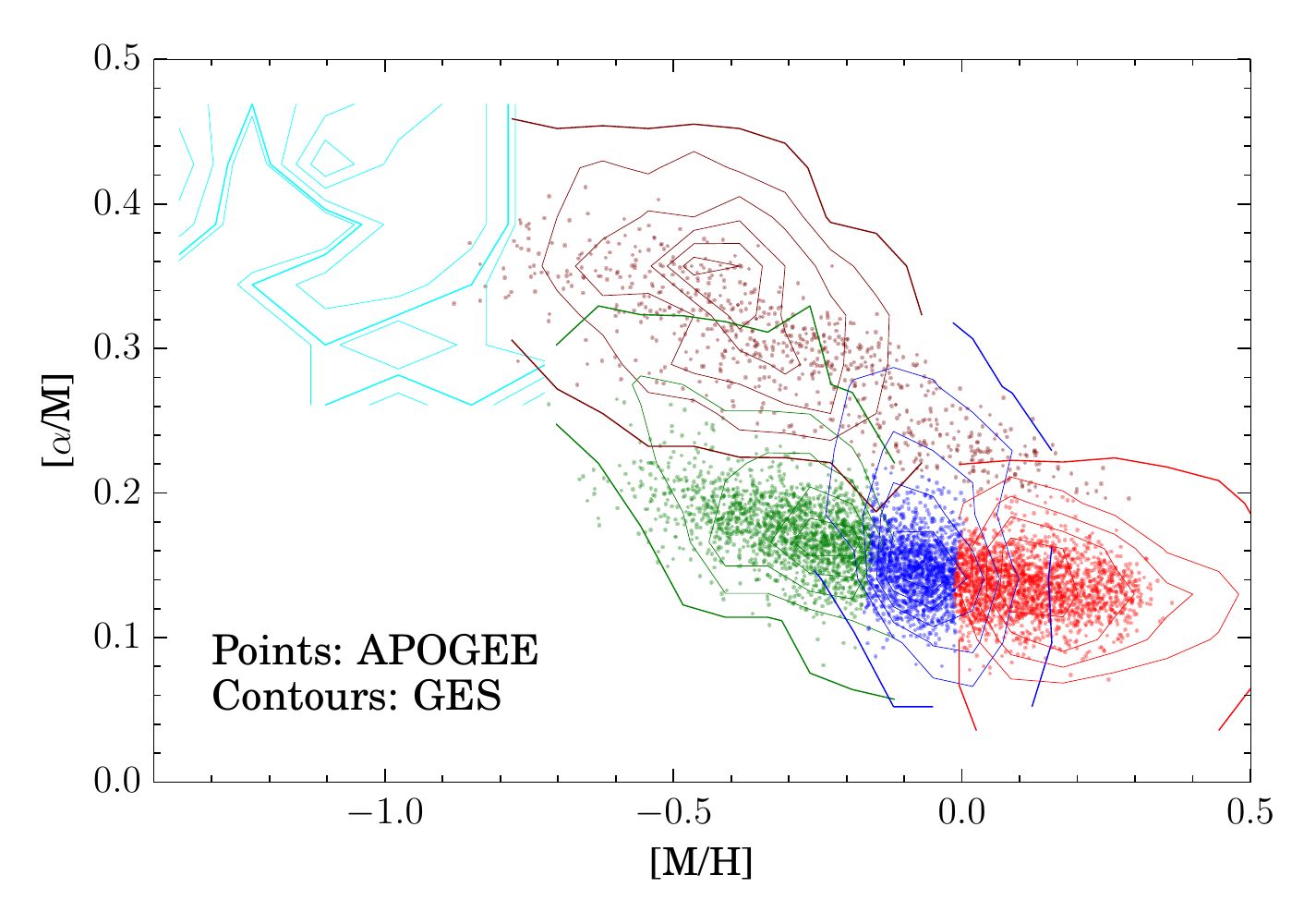}
\caption{GMM decomposition of Gaia-ESO survey and APOGEE samples in the [$\alpha$/M] vs. [M/H] plane. Points depict the APOGEE \textit{\textup{working sample}}. Contour lines draw the density distribution of Gaia-ESO survey GMM data groups. A vertical shift of $\Delta\textmd{[}\alpha/\textmd{M]}=0.1$ dex was applied to the APOGEE sample to obtain a better agreement between the two data sets. Color coding is set to be consistent with the one used to represent data groups throughout this paper.} 
\label{fig:gmm_comparacion_apogee}
\end{center}
\end{figure}

The resulting GMM data group distributions are  displayed with color points in Fig. \ref{fig:gmm_comparacion_apogee}. As in Fig. \ref{fig:gmm_vs_split_manual} for the GES sample, the APOGEE sample is separated into subgroups associated with the thin and thick disk sequences. Since the selection function of APOGEE is biased against metal-poor stars ([Fe/H ]<-0.9 dex, given the survey color cut at $\textmd{(J-K)}_0\geq 0.5$, \citeauthor{zasowski2013} \citeyear{zasowski2013}), there are no data points in the halo region of the abundance-metallicity plane, and consequently, there is no GMM component accounting for them as in the Gaia-ESO sample. The centroids and marginalized dispersions of the resulting model components are listed in the rightmost columns of Table \ref{tab:gmm_results}.

To compare our results with APOGEE, we computed contour plots from the [M/H] and [$\alpha$/Fe] values provided in GES data. As pointed out in Sect. \ref{subsec:apogee_final_sample}, for APOGEE, [M/H] and [$\alpha$/M] are part of the set of recommended fundamental parameters and are calibrated with respect to external  spectroscopic references. We verified that accounting for errors in data, [M/H] correlates quite well with [Fe/H], so that [$\alpha$/Fe] (the abundance ratio we have in GES) is a good proxy for [$\alpha$/M]. To compute GES contour plots, we used the GMM classification in the [Mg/Fe] vs. [Fe/H] plane (Fig. \ref{fig:gmm_vs_split_manual}) to separate the sample into the different sequences. After assigning group membership labels to the stars in the sample, we considered them in the [$\alpha$/Fe] vs. [M/H] plane. We used the data distribution on this plane  to calculate contours delineating their density distribution. We overplot them with the APOGEE data as color lines in Fig. \ref{fig:gmm_comparacion_apogee}. This figure shows that the qualitative agreement is quite satisfactory after applying a vertical shift of $\Delta\textmd{[}\alpha/\textmd{M]}=0.1$ dex to the APOGEE sample. The difference is probably caused by the different chemical calibration strategy or the different definition of global metallicity adopted by the two surveys. The dispersion of GES contours is larger (but contours delineating high-density regions are tight to the APOGEE distribution) than that characteristic of APOGEE data, but the general agreement is very good. In particular, the thin disk is separated into three subgroups in both cases, with qualitatively comparable shapes.

We recall that the Gaia-ESO and APOGEE surveys constitute two different and independent observational approaches. They have a different selection function, the observations are carried with different instruments, spectral ranges, resolutions, parameter determination, and calibrations. In spite of these conceptual differences, the distributions in the abundance-metallicity plane are qualitatively comparable, which demonstrates that possible observational biases do not create hindering artifacts in the sample distributions. In conclusion, we can claim that the definition of the different sequences, halo and disks, with their associated substructures as defined by the GES sample, are confirmed from APOGEE data.

\begin{table*}
\caption{Centroids and marginalized dispersions of the GMM modes for different samples. The analysis was performed in the [Mg/Fe] vs. [Fe/H] plane for the GES samples and in the [$\alpha$/M] vs [M/H] for APOGEE.}
\begin{flushleft}
\begin{tabular}{ccccccccc}
\noalign{\smallskip}
\hline
\noalign{\smallskip}

Group  & \multispan2 GES-All & \multispan2 GES-RC & \multispan2 GES-Dwarfs  & \multispan2 APOGEE \\
\cmidrule(r){2-3} \cmidrule(r){4-5} \cmidrule(r){6-7} \cmidrule(r){8-9} \ 
 &  $\mu$    & $\sigma$ &   $\mu$ &   $\sigma$  & $\mu$ & $\sigma$ & $\mu$ & $\sigma$ \\
 
\noalign{\smallskip} 
\hline 
\noalign{\smallskip} 

Cyan  &  (-1.09, 0.38) & (0.23, 0.09)  &   (-1.19, 0.39) & (0.22, 0.09)  &   (-0.50, 0.20) & (0.27, 0.12)  &   (---, ---) & (---,---)        \\
Brown &  (-0.49, 0.30) & (0.18, 0.07)  &   (-0.49, 0.33) & (0.17, 0.07)  &   (-0.49, 0.28) & (0.18, 0.07)  &   (-0.29, 0.19) & (0.24, 0.05)  \\
Green &  (-0.35, 0.09) & (0.21, 0.08)  &   (-0.41, 0.15) & (0.12, 0.07)  &   (-0.26, 0.07) & (0.21, 0.08)  &   (-0.26, 0.07) & (0.14, 0.03)  \\
Blue  &  (-0.12, 0.07) & (0.13, 0.09)  &   (-0.13, 0.11) & (0.07, 0.07)  &   (-0.10, 0.04) & (0.13, 0.09)  &   (-0.08, 0.05) & (0.08, 0.03)  \\
Red   &  (0.11,  0.00) & (0.17, 0.06)  &   (0.19,  0.01) & (0.17, 0.06)  &   (0.10,  0.00) & (0.14, 0.06)  &   ( 0.11, 0.04) & (0.10, 0.02)  \\

\noalign{\smallskip} \hline \end{tabular} 
\label{tab:gmm_results} 
\end{flushleft} 
\end{table*} 

%==================================================================================================
\subsection{GMM results for different stellar luminosity classes}
\label{subsec:gmm_data_groups}
\begin{figure}
\begin{center}
\includegraphics[width=9cm]{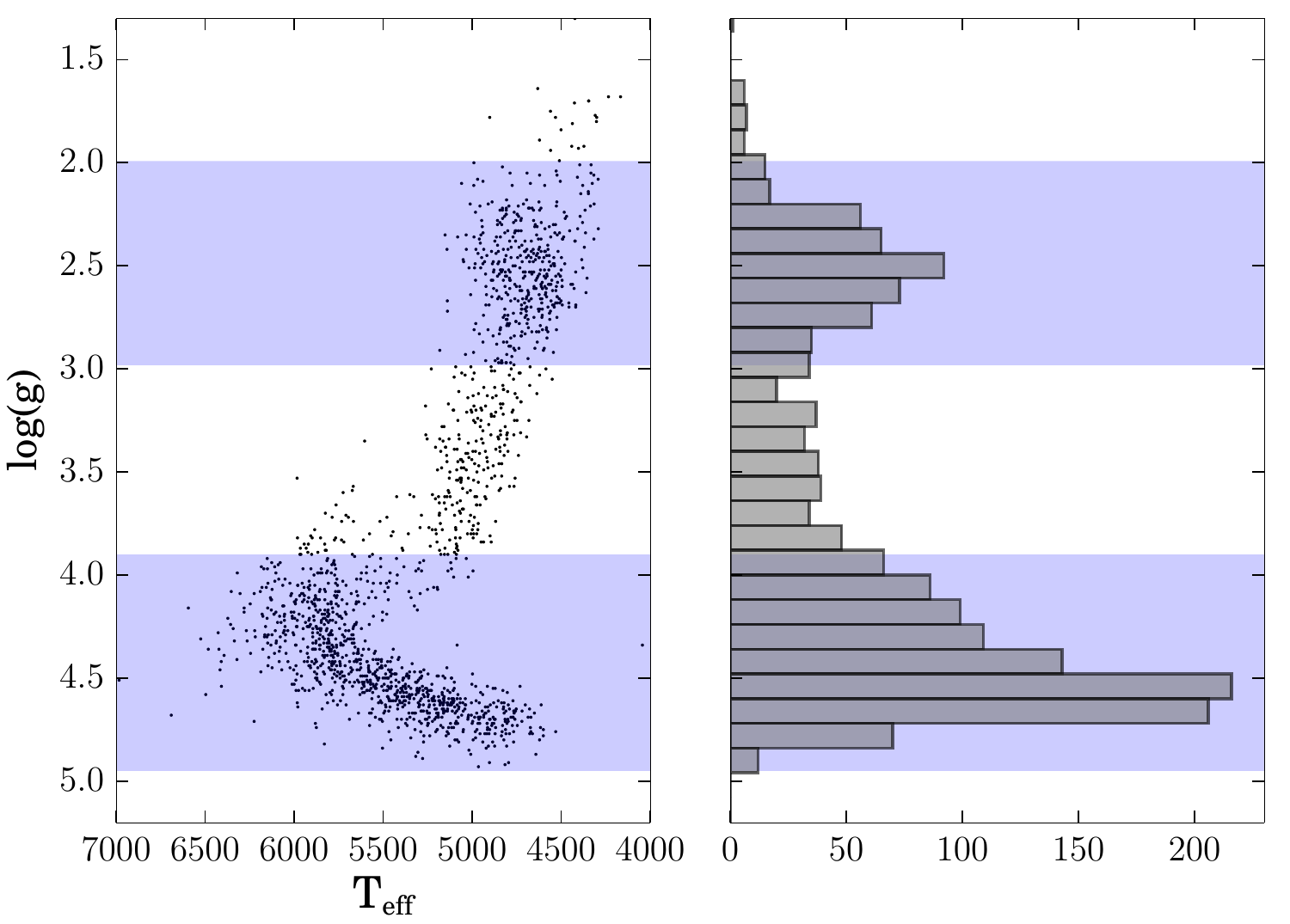}
\caption{\textit{Left:} HR diagram for the complete sample with shaded bands highlighting the RC and dwarf regions. \textit{Right:} histogram of the $\log(g)$ distribution. The RC region is defined as the stars located between 2.0 and 3.0 dex. The dwarf region is defined by $\log(g)>3.9$ dex.} 
\label{fig:rc_splitting}
\end{center}
\end{figure}

As shown in the left panel of Fig. \ref{fig:rc_splitting}, the GES sample covers all the main stellar evolutionary stages of low-mass stars. We wish to determine whether the GMM results are affected by the GES selection function given the different evolutionary stages selected in the sample. Figure \ref{fig:rc_splitting} clearly shows that most of the stars are in the dwarf sequence or in the RC region. The distribution of \logg values in the right panel exhibits two main peaks, one accounting for the RC, and the other for the dwarf sequence. In the following, we adopt as the RC region the one spanning \logg values between 2.0 and 3.0 dex and as the dwarf region that inhabited by stars with \logg$>3.9$ dex.

\begin{figure}
\begin{center}
\includegraphics[width=8.8cm]{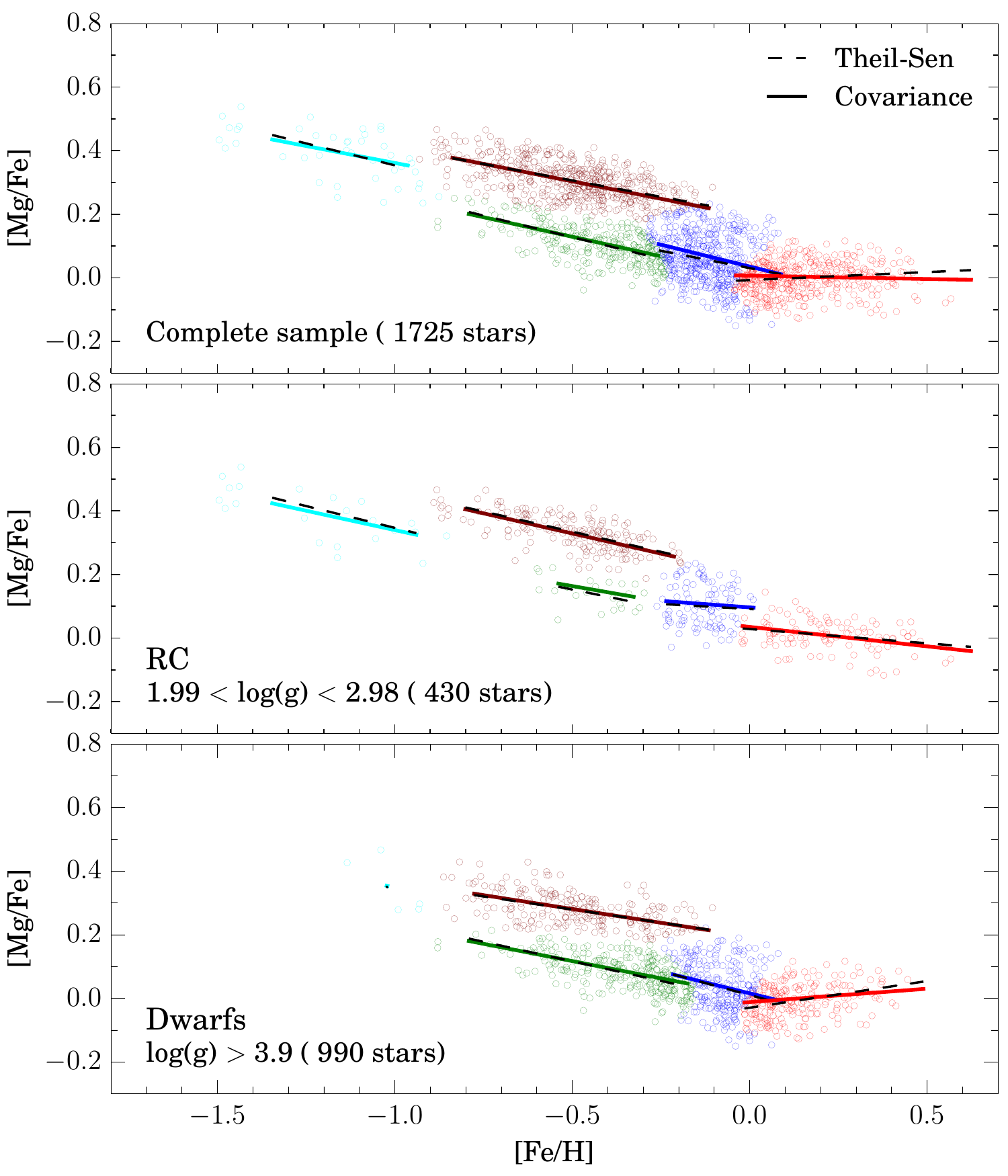}
\caption{GMM decomposition performed on the complete GES sample (\textit{upper panel}), RC stars (\textit{middle panel}), and dwarfs (\textit{lower panel}). Solid black lines represent trend lines determined for each GMM group from the mean and covariance of the respective mode of the mixture. Dashed black lines stand for trends computed directly from the datapoints, using the Theil-Sen estimator.} 
\label{fig:gmm_rc_vs_enanas}
\end{center}
\end{figure}

To perform our test, we considered two subsamples, one in the RC region (430 stars) and another in the dwarf sequence (990 stars). The resulting GMM decompositions in the plane [Mg/Fe] vs. [Fe/H] are shown in the lower panels of Fig. \ref{fig:gmm_rc_vs_enanas}. The results for the two subsamples agree qualitatively well with respect to the total sample.

In particular, we can conclude from Fig. \ref{fig:gmm_rc_vs_enanas} that\begin{itemize}
\item RC and dwarf samples present the same qualitative data substructure, as highlighted by the respective GMM decompositions.

\item The RC sample presents indications for a number underdensity in the thin disk sequence at [Fe/H]$\sim-0.25$ and [Fe/H]$\sim0.05$ dex .

\item The sequences drawn by RC stars seem to be less dispersed in [Mg/Fe] than those corresponding to dwarf stars: from comparing the different GMM groups, we see that the dispersion in RC stars (around the linear fits of the GMM groups) is around 10\% smaller than in dwarfs.
\end{itemize}

For the last point, we can verify whether the effect is due to larger errors in the abundances derived for dwarf stars or if it corresponds to a different behavior of the two populations. An examination of the error distributions of [Mg/Fe] and [Fe/H] in both groups of stars shows that, in general, RC stars have larger errors than dwarfs (i.e., mean errors of $\overline{\textmd{E[Fe/H]}}=0.07$ dex; $\overline{\textmd{E[Mg/Fe]}}=0.09$ dex for RC, and $\overline{\textmd{E[Fe/H]}}=0.05$ dex; $\overline{\textmd{E[Mg/Fe]}}=0.07$ dex for dwarfs), in
contrast to the smaller dispersion in [Mg/Fe] observed in this group. As the errors in [Mg/Fe] are of the same order of magnitude as the observed dispersions around the linear fits, no conclusions can be drawn with respect to a possible astrophysical difference in the dispersion of RC and dwarf stars.

Using  the covariance information of the GMM results, we  estimated the linear trends of each group for the three samples shown in Fig. \ref{fig:gmm_rc_vs_enanas}. They are depicted as solid color lines. We also estimated linear fits directly from the data in each family group, using the Theil-Sen estimator. For these last estimates, we computed errors by bootstrapping. To do this, we resampled the respective data groups to generate 500 bootstrap samples. We estimated the slope for each resampling. The error is taken as their standard deviation. The results for the three panels are presented in Table \ref{tab:pends_grupos}. These slopes, in particular for the subsolar metallicity part of the thin disk (green and blue groups), compare well to those determined by \citet{kordopatis2015b} (cf. their Table 1).

\begin{table} 
\centering
\caption{Slopes for the line models fitted to the data in Fig. \ref{fig:gmm_rc_vs_enanas}. The slope is estimated using the Theil-Sen estimator from the complete dataset (Col. 2), the RC stars alone (Col. 3), and dwarfs (Col. 4). Reported errors are estimated by bootstrapping.}
\begin{tabular}{lccc}
\hline
\hline
Group   &     m (Total)      &   m (RC)         &  m (Dwarfs)    \\\hline
Cyan    &  -0.27$\pm$0.064 &  -0.27$\pm$0.059 &  -0.88$\pm$1.147 \\
Brown  &  -0.21$\pm$0.014 &  -0.25$\pm$0.020 &  -0.16$\pm$0.022 \\
Green   &  -0.28$\pm$0.020 &  -0.23$\pm$0.110 &  -0.25$\pm$0.019 \\
Blue    &  -0.23$\pm$0.054 &  -0.06$\pm$0.115 &  -0.29$\pm$0.095 \\
Red     &   0.05$\pm$0.018 &  -0.09$\pm$0.035 &   0.17$\pm$0.032 \\\hline\end{tabular}
\label{tab:pends_grupos}
\end{table}

In the thin disk sequence of the complete sample, a break in slope is visible at [Fe/H]$\sim$0.0 dex, at the location of the boundary between the two metal-rich components. This is equally visible in the dwarf subsample, but less so in the RC subsample. We computed the Student t-test for the slope difference significance of the complete sample, obtaining $t(912)=-6.615$ with an associated $p$-value$\sim0$, thus confirming its statistical significance. A similar result is obtained considering only the dwarf sample, with a $p$-value$\sim0$. For the RC distribution, the same analysis gives us a $p$-value=0.39, showing that, in this case, the difference between the slopes is not statistically significant. 
In particular, while comparing the two subsamples, the metal-rich group presents a negative slope for RC, but a positive one for dwarf stars (Fig. \ref{fig:gmm_rc_vs_enanas} and Table \ref{tab:pends_grupos}). 

These slope differences can be explained as an effect of comparing stars with distinct stellar parameters, since the ranges in \teff and \logg of RC and dwarfs are different. Differences in the atmospheric models and line formation modeling at the respective physical regimes can lead us to slightly different results by including different systematics. We checked for this possibility by constructing explicit plots between [Mg/Fe] and atmospheric parameters. No clear correlations were found. Still, some age differences (that cannot be verified here) can imply differences in the chemical enrichment patterns. A more plausible explanation for this effect comes from the distribution of RC and dwarf stars in our sample. RC stars mostly sample regions inside the solar annulus at R$_{\textmd GC}<7$ kpc, while dwarfs span a smaller range around the solar position. In this sense, the observed slope differences could arise as an effect of different chemical evolution with R$_{\textmd GC}$.

Another interesting feature in Fig. \ref{fig:gmm_rc_vs_enanas} is the density gap that appears in the RC sequence at solar metallicity, where the dwarf sample presents an overdensity. A plausible explanation could be related with the GES selection function. RC stars are indeed intrinsically bright, falling out of the magnitude limits of the photometric selection when they are close. These excluded stars are precisely those that show a metallicity that is characteristic of the solar neighborhood. As a consequence, they remain undersampled in the respective distribution in the [Mg/Fe] vs. [Fe/H] plane. 

On the other hand, as our RC stars are mainly sampling internal regions of the disk, the thin disk stars should preferentially be metal-rich members. This can explain the scarcity of low-$\alpha$ stars observed at [Fe/H]$\sim-0.25$ dex in the middle panel of Fig. \ref{fig:gmm_rc_vs_enanas}.

Overall, we can conclude that our GMM decomposition in the [Mg/Fe] vs. [Fe/H] is essentially the same when RC and dwarfs subsamples are compared. This demonstrates that chemical patterns depend only on the spatial distribution of the sampled population, with no bias from the GES selection function.

%==================================================================================================
%   The metal-poor thin disk
%==================================================================================================
\section{The metal-poor thin disk}
\label{sec:metal-poor_thin_disk}
Given the different sequences identified in the abundance-metallicity plane, we explored their spatial and kinematic distributions
to compare them. This approach might provide interesting insights into understanding their comparatively different origins and/or chemodynamical evolution. We focused particularly on the metal-poor thin disk GMM subgroup at [Fe/H]$<-0.25$ dex (green points in Fig. \ref{fig:gmm_vs_split_manual}). It separates stars for which recent studies claimed a possible evolution disconnected from the one of the thick and metal-rich thin disk \citep{haywood2013,snaith2015}.

\begin{figure}
\begin{center}
\includegraphics[width=8.7cm]{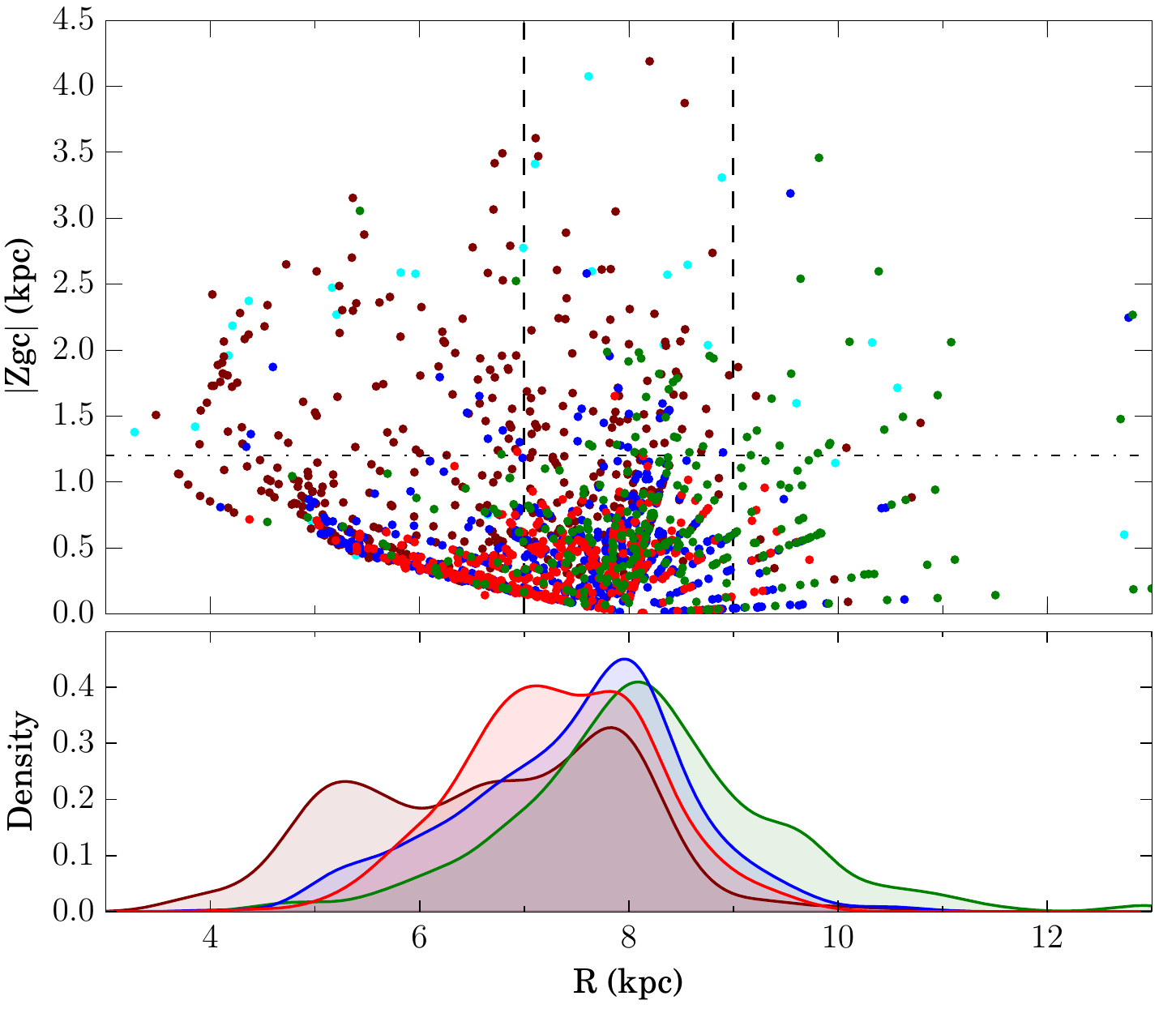}
\caption{\textit{Upper panel: }Distribution of the  sample in the $|Z|$ vs. R plane. The points are colored as in Fig \ref{fig:gmm_vs_split_manual}. Two vertical dashed lines delimit the solar region $7<\textmd{R}<9$ kpc, while a point-dashed horizontal line is drawn at $|Z|=1.2$ kpc. \textit{Lower panel: } Normalized generalized histograms (kernel 0.5 kpc) showing the radial distribution of the main GMM data groups (excluding the halo stars; cyan group). They are computed using only stars with $|Z|<1.2$ kpc to reduce the effect of undersampled regions far from the Galactic plane.} 
\label{fig:r_vs_z_restricciones}
\end{center}
\end{figure}

Figure \ref{fig:r_vs_z_restricciones} displays the general distribution of our sample in the $|\textmd{Z}|$ vs. R$_{\textmd GC}$ plane. As usual, points and profiles are colored according to the GMM groups. This figure shows a tendency for the metal-poor thin disk  stars to be more distributed toward the outer regions (as seen also in other works, e.g., \citeauthor{bovy2012b} \citeyear{bovy2012b}, \citeauthor{anders2014} \citeyear{anders2014}). The metal-rich thin disk group presents an excess of stars toward inner regions (which looks compatible with predictions of the MCM model \citeauthor{minchev2013} \citeyear{minchev2013}, e.g., their Figure 3). Finally, the thick disk group shows a distribution extending well into the inner regions, but also with important contribution at the solar radii, although clearly less important outside the solar circle (as suggested also by the first-year APOGEE data; \citeauthor{anders2014} \citeyear{anders2014}).

\begin{figure*}
\begin{center}
\includegraphics[width=6.1cm]{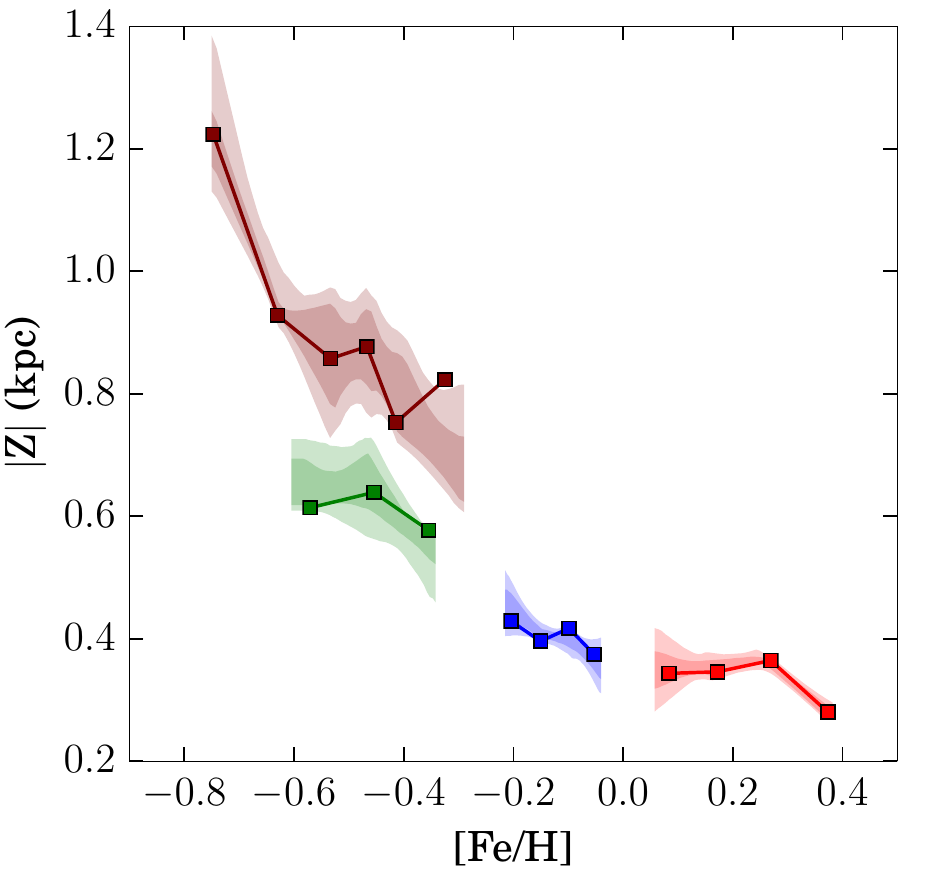}
\includegraphics[width=6.1cm]{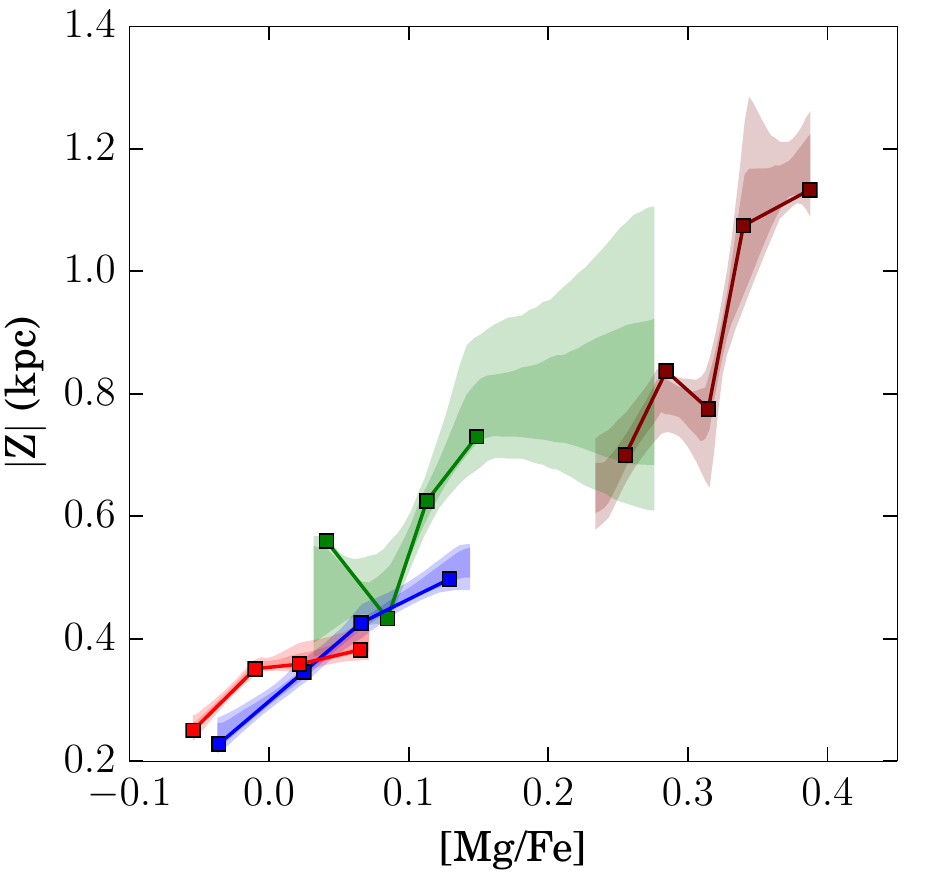}\\
\caption{Vertical distribution of disk stars as a function of [Fe/H] and [Mg/Fe]. Colors according to the GMM groups as in Fig \ref{fig:gmm_vs_split_manual}. The curve+points correspond to the median values of stars separated in bins equally populated. Error bands at the $1\sigma$ and $2\sigma$ level are computed by Monte Carlo resamplings to take into account the effect of errors in metallicity and abundances on the sample classification.} 
\label{fig:distros_z}
\end{center}
\end{figure*}

\subsection{Vertical distribution of disk stars}
\label{subsec:vertical_distro}

It is interesting to check how the disk GMM groups in Fig. \ref{fig:gmm_vs_split_manual} are distributed in distance with respect to the plane. To do this, we used their absolute values  $|\textmd{Z}|$.

Figure \ref{fig:r_vs_z_restricciones} shows the general predominance of thick disk and halo group stars at larger distances from the plane. On the other hand, the relative distribution of thin disk stars is unclear since they crowd close to the plane. To obtain a cleaner picture, we can calculate a central tendency estimate to summarize the behavior of $|\textmd{Z}|$ with respect to some other property such as [Fe/H] or [Mg/Fe].

Before doing so, we can stop to consider what our GMM sample classification imposes on our data set. As explained in Sect. \ref{sec:decomp_ges_sample}, the data classification into GMM groups (depicted by the different colors in Fig. \ref{fig:gmm_vs_split_manual}) was  performed by taking for each star the maximum of the set of responsibilities for this observation to be explained by each one of the five GMM components. This simple labeling process does not take into account the fact that some points, because of their position in the abundance-metallicity plane, are intrinsically prone to belong with similar probabilities  to two or even more components. Even if the errors in [Fe/H] and [Mg/Fe] are very small, these points will have a higher level of uncertainty in their classification. In order to study the distribution of physical properties of the GMM groups, $|\textmd{Z}|$  here, for example, it is important to ensure that we work with a sample whose members are strongly explained by the components they are classified in.
We performed 2000 Monte Carlo resamplings of the complete sample by generating repetitions for each star from their error ellipsoid in [Fe/H] and [Mg/Fe]\footnote{The primary quantities and errors determined in iDR2 are [Fe/H] and [Mg/H]. This implies error correlations in the Monte Carlo realizations in the [Mg/Fe] vs. [Fe/H] plane, which we took into account.}. For each resampling, we computed the responsibility vector for each star. Since the individual position of the stars change in the abundance-metallicity plane in each repetition, their set of responsibilities changes
as well. From the set of resamplings, we computed for each star the set of mean responsibilities and their associated dispersions. For a given star, if  $P_1$ and $\sigma_1$  are the highest responsibility and its dispersion, and $P_2$ the second highest responsibility, then
$$|P_1-P_2|>N\sigma_1$$
defines a criterion that allows us to keep stars whose highest responsibility is at a statistically significant distance from the second highest responsibility.

We adopted $N=1.2$ to define a statistical subsample composed of 1252 stars (from the initial GES sample of 1725 stars). As expected, this criterion removes stars located in [Fe/H] and [Mg/Fe] plane regions where GMM groups intersect.

In Fig. \ref{fig:distros_z} we use the statistical subsample to make explicit plots of $|\textmd{Z}|$ as a function of [Fe/H] and [Mg/Fe]. For each GMM data group, we computed the median of $|\textmd{Z}|$ values in equally populated bins of [Fe/H] or [Mg/Fe] (with a minimum of 40 for the green and a maximum of 90 datapoints for the blue groups). The resulting curves are displayed  using square symbols and the same colors as in Fig. \ref{fig:gmm_vs_split_manual}. 

Given the errors in [Fe/H] and [Mg/Fe] and our highest responsibility-based labeling criterion, our sample can be affected by some degree of missclassification. We can estimate its effect in the curves of Fig. \ref{fig:distros_z}. To do this, we computed 2000 Monte Carlo resamplings of the stars in the [Mg/Fe] vs. [Fe/H] plane. For each repetition, we classified the sample into the GMM groups defined by the original sample and computed the median curves for each component. From the complete set of curves for each GMM component, we  determined deviations from the median trends. The shaded  error bands in Fig.\ref{fig:distros_z} represent the deviations at the $1\sigma$ and $2\sigma$ level from the median tendency of the Monte Carlo resampled curves.

The left panel of Fig. \ref{fig:distros_z} shows that the metal-intermediate and metal-rich groups of the thin disk (blue and red components) present a monotonous slight decrease of the mean $|\textmd{Z}|$ with metallicity. In addition, the mean $|\textmd{Z}|$ values of the metal poor thin disk (green) are on average 0.2 kpc higher than the more metal-rich thin disk. In this sense, the metal-poor thin disk appears to be of intermediate thickness of the thick and metal-rich thin disks (although the measurement of the relative scale heights is beyond the scope of this paper).

If we use [Mg/Fe] as a proxy for the age of thick disk stars (this is justified by the tight correlation between the two found for instance by \citeauthor{haywood2013} \citeyear{haywood2013}), we can see in the right panel of Fig. \ref{fig:distros_z} a decline in the thick disk thickness with age. Older stars are distributed at larger distances from the plane. 

In spite of the less clear correlation between age and [Mg/Fe] for thin disk stars in \citet{haywood2013}, similar trends with $|\textmd{Z}|$ can be suggested for the thin disk sequence. The most likely oldest (Mg-enhanced) stars of the metal-poor thin disk present mean $|\textmd{Z}|$ values that are comparable to the younger part of the thick disk. The two GMM families standing for the metal-intermediate and  metal-rich thin disk define a continuous decrement in thickness with age. Although the metal-rich thin disk stars are a continuation of the metal-intermediate stars in the $|\textmd{Z}|$ vs. metallicity plane, they are not in the $|\textmd{Z}|$ vs. [Mg/Fe] one. Indeed, it has been found that many of the solar neighborhood metal-rich stars are in fact old \citep{trevisan2011}.

The above results remain valid if instead of using the total sample, we restrict it to solar neighborhood members at $7<\textmd{R}<9$ kpc,  the region depicted by the two vertical dashed lines in the upper panel of Fig. \ref{fig:r_vs_z_restricciones}. The size of our sample does not allow us to trace radial variations in the properties described here.

These qualitative comparisons are in relative terms and do not aim to provide an absolute idea of the scale height of the disk components.

In conclusion, from a statistically well-defined sample of stars classified into the GMM groups,  we find that the metal-poor thin disk presents a characteristic thickness intermediate between those of the metal-rich thin and thick disks. The effects of star misclassification, given the errors on [Fe/H] and [Mg/Fe] of the stars in the sample, are small, enforcing the statistical significance of the reported differences.

%==================================================================================================
\subsection{Rotational velocity of disk stars}
\label{subsec:rotational_vel}
Kinematical data were adopted from \citet{guiglion2015}, where cylindrical velocity components were calculated for a larger subsample of $\sim$6800 stars from the  Gaia-ESO iDR2. We refer to this paper for further details.

As a first characterization of our sample, we checked the density distributions of V$_\phi$ for the GMM families. Narrow distributions with peaks at V$_\phi\sim210$ \kms stand for the three thin disk GMM groups, while a broader distribution centered at lower V$_\phi\sim180$ \kms is found for the thick disk stars. The metal-poor group attributed to the halo presents a very broad distribution of values around zero, compatible with their expected spheroidal kinematics.

\begin{figure}
\begin{center}
\includegraphics[width=8cm]{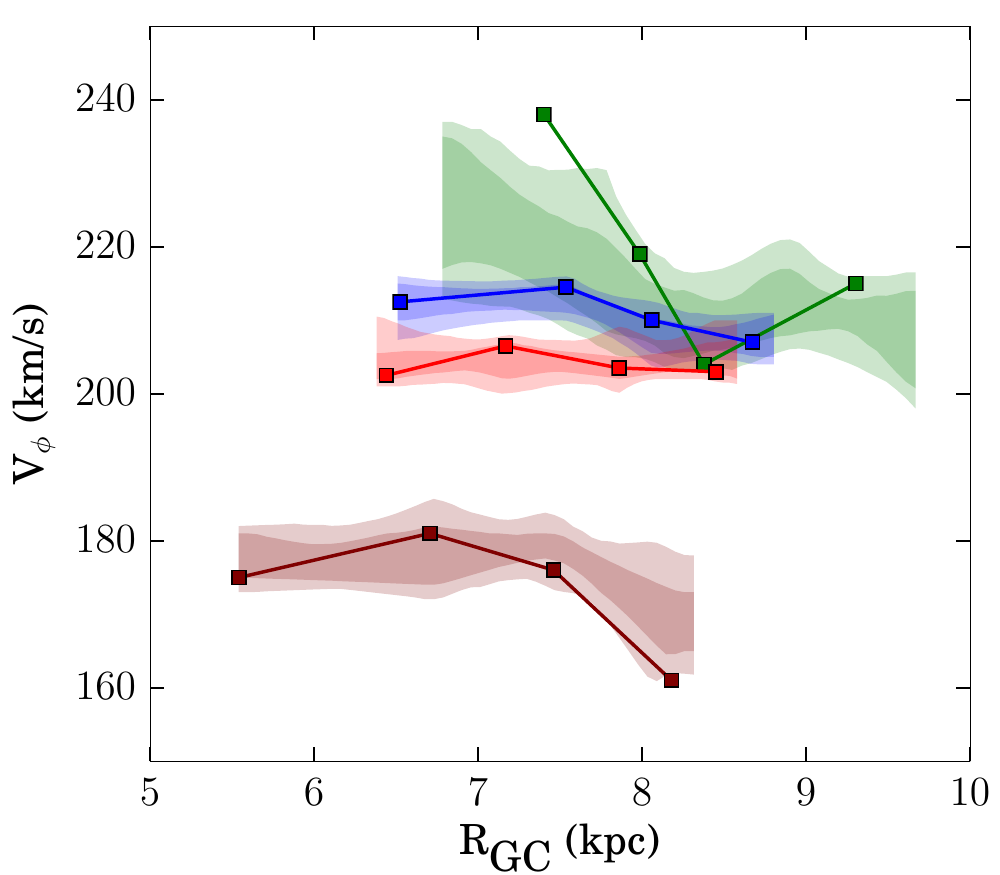}
\caption{Azimuthal velocity component V$_{\phi}$ as a function of the Galactocentric radius R$_{\textmd GC}$. Colors as in Fig. \ref{fig:gmm_vs_split_manual}. For each GMM data family, a central tendency curve is constructed by separating the dataset into a number of equally populated bins in R$_{\textmd GC}$. Median values inside each bin determine the position of the dots in the curves. Only stars with $5<{\textmd R}_{\textmd GC}<10$ kpc are considered to avoid undersampled bins in the extremes of the distribution. Error bands are computed at the $1\sigma$ and $2\sigma$ level by Monte Carlo resamplings to take into account the effect of errors in metallicity and abundances on the sample classification.}
\label{fig:r_vs_vphi}
\end{center}
\end{figure}

In Fig. \ref{fig:r_vs_vphi} we display the azimuthal velocity component V$_\phi$ as a function of the cylindrical Galactocentric radius R$_{\textmd GC}$. Only stars from the statistical subsample defined in Sect. \ref{subsec:vertical_distro}  with $5<{\textmd R}_{\textmd GC}<10$ kpc were used to avoid undersampling outside these limits (see Fig. \ref{fig:r_vs_z_restricciones}). The four curves stand for the median tendency of the stars in each GMM data group. Error bands are computed in the same way as in Fig. \ref{fig:distros_z} to take into account the effects of sample misclassification.

The analysis and interpretation of kinematical data can be complicated by the fact that errors in stellar parameters, distances, and proper motions propagate to the computed Galactocentric velocities, eventually introducing strong outliers. The position of points defining curves in Fig. \ref{fig:r_vs_vphi} corresponds to the median of stars separated in equally populated bins in R$_{\textmd GC}$\footnote{Bin populations are: brown 70 points, green 33 points, blue 90 points and red 70 points.}. The use of medians is convenient in this case to take advantage of its robustness with respect to strong influential outliers. To further ensure the robustness of the observed patterns, we cleaned the sample by removing stars with an accuracy in V$_\phi$ lower than a given cutoff. Different values were tried in consideration of the error distribution of V$_\phi$. We verified that no differences were introduced by allowing points with higher inaccuracies in the sample. In Fig. \ref{fig:r_vs_vphi} we use stars with accuracies better than 90 \kms, thereby removing completely the tail of high-inaccuracy stars without reducing the sample size too much. Most of the eliminated stars belong to the thick disk. After all these cuts, the statistical subsample used in Fig. \ref{fig:r_vs_vphi}  is reduced to 1026 stars.

Different tendencies are observed for the azimuthal velocity of the GMM components. A decreasing profile out of R$_\textmd{GC}\sim7$ kpc is observed for thick disk stars, in agreement with \citet[][ their Fig. 12]{guiglion2015}. The two metal-intermediate and  metal-rich  (blue and red) thin disk groups display nearly flat profiles centered at 210 \kms. A distinct pattern is displayed by the metal-poor thin disk group of stars (green), with a peak at V$_\phi\sim235$ \kms for stars in the range $\sim$7-8 kpc. This bin contains stars in the metal-rich side of the group with [Fe/H] distributed mostly between -0.3 and -0.5 dex. A decrease in V$_\phi$ is observed for this group at distances R$_{\textmd GC}>8$ kpc, where the stars  span a wider metallicity range to [Fe/H]$\sim-0.8$ dex.

If we consider the current Galactocentric radius distributions in Fig. \ref{fig:r_vs_z_restricciones} as a blurred version of the distribution of their guiding radii, we can figure out a possible explanation for the pattern observed in Fig. \ref{fig:r_vs_vphi}. The metal-rich group stars (red) are more dominated by stars with inner guiding radii and hence display a slower rotation profile. The metal-intermediate group stars (blue) are more concentrated on the solar radius and hence are closer on average to the solar value of 220 \kms. On the other hand, the metal-poor group has more stars that would have guiding radii in the outer regions. The fact that they appear at smaller current Galactocentric distances can be due to orbital blurring in the sense of epicyclic motion, or because their orbits are eccentric. In this sense, stars at a current distance close to the Sun will have higher azimuthal velocities as we found here.

The aspect of the error bands shows that in general misclassification does not significantly condition our results. A larger deviation of the error-band profile with respect to the median tendency of the sample is observed in the metal-poor thin disk group. This effect can be explained by stars that because of their relatively larger errors in the [Mg/Fe] vs. [Fe/H] plane, change of classification between thick disk and metal-poor thin disk during the Monte Carlo resamplings. Stars with thick disk kinematics being classified as metal-poor thin disk can pull down its median V$_\phi$ values. This effect is expected to be stronger for distant stars because of their a priori larger errors in abundance and metallicity. In spite of this effect, the median tendency in V$_\phi$ of the Monte Carlo resamplings of the metal-poor thin disk can still be separated from those of the metal-rich thin disk groups in the region R$_{\textmd GC}<8$ kpc. This enforces the significance of the differences found between the statistical sample profiles of these groups.

Complementing information from Figs. \ref{fig:distros_z} and \ref{fig:r_vs_vphi} can be used to further highlight the differences between the metal-rich and metal-poor regions of the thin disk. Below we discuss this in the context of some other recent results.

%==================================================================================================
%    Discussion
%==================================================================================================
\section{Discussion}
\label{sec:discusion}

The study of the disk(s) abundance patterns and their spatial and temporal dependency is one of the pathways to understand the formation history and subsequent evolution of the structures assembled in the Galactic disk system.
A chemical classification of the disk stars from their distribution in the plane [$\alpha$/Fe] vs. [Fe/H] has proven to be a fruitful way to separate clean samples of thin and thick disk stars with less superposition in other properties.

Our GMM analysis allows us to separate the selected Gaia-ESO iDR2 subsample in a way compatible with other approaches that separated thin  and thick disk sequences by following the minimum density in [Mg/Fe] in several metallicity bins \citep[as in][]{recio-blanco2014,mikolaitis2014}, or used a pure maximum likelihood procedure in narrow metallicity bins \citep{kordopatis2015b}. In addition, our approach reveals some number density and slope variations of the thin disk track in the abundance-metallicity plane.

In this sense, if we consider the metal-rich and metal-intermediate part of the thin disk sequence (two of the three GMM groups into which the thin disk sequence is separated; $\textmd{[Fe/H]>-0.25}$ dex), with stars located preferentially at R$_{\textmd GC}<8.5$ kpc (Fig. \ref{fig:r_vs_z_restricciones}), we see that they constitute a sequence with a slope break point at solar metallicity. The qualitative similarity between RC and dwarf distributions (Fig. \ref{fig:gmm_rc_vs_enanas}) shows that this slope variation is present at different Galactocentric radial distances. These observations suggest that the chemical enrichment history of the thin disk might present some critical transition at solar metallicity.

A comprehensive analysis of the possible radial variations of the slopes and breaking point for the metal-rich thin disk stars should be based on a larger data set. Unfortunately, our sample size does not allow us to perform such an analysis with enough statistical significance. This will be attempted using the APOGEE data presented in Sect. \ref{subsec:gmm_ges_data_sample} in a future publication.

In addition to the metal-intermediate and metal-rich thin disk groups, our cluster analysis led us to further separate a group at its metal-poor end, with $[\textmd{Fe/H]}<-0.25$ dex. This region of the [$\alpha$/Fe] vs. [Fe/H] distribution has been the object of  recent debate concerning some particular properties that place it in a transition status between the thin and thick disks \citep{haywood2013}. Our data (Fig. \ref{fig:r_vs_z_restricciones}, lower panel) show that the radial distribution of these stars is shifted to larger radial distances with respect to those of the thick, the metal-intermediate, and metal-rich thin disk groups. From the more radially extended APOGEE DR12 sample of \citet{hayden2015} \citep[also  ][]{anders2014} it is possible to see that $\alpha$-poor stars with [Fe/H]$\sim-0.4$ to $-0.5$ dex are widely dominant in number at larger Galactocentric distances (their Fig. 4). The spatial distribution of metal-poor thin disk stars with respect to the plane distinguishes them from the more metal-rich internal parts, since the characteristic $|\textmd{Z}|$ median values are in general higher over the whole abundance and metallicity extent (Fig. \ref{fig:distros_z}). Their azimuthal velocity distribution in Fig. \ref{fig:r_vs_vphi} display some complex pattern; while its stars at $7\leq\textmd{R}_{\textmd GC}\leq 8$ kpc are markedly different, with a V$_\phi$ peak of $\sim235$ \kms, the more external R$_{\textmd GC}>8$ kpc stars display slower rotation. In an inside-out formation scenario for the thin disk, these properties of the metal-poor group could be explained by considering them as a more warped distribution of the younger populations toward the outer radii. In fact, simulations predict that coeval populations always flare \citep{minchev2015}.  If the outer metal-poor thin disk is build up from the superposition of  young populations flaring at larger Galactocentric distances, its scale height will naturally be larger than the one of the stars in the inner disk. The fact that these stars are distributed at larger distances from the plane could then also explain their smaller azimuthal velocities at large R$_{\textmd GC}$.

While this is a plausible explanation, some evidence indicates that the stars in the metal-poor end of the thin disk are in fact old \citep{haywood2013}, which is qualitatively supported by the on average higher Mg enhancements compared with the more metal-rich portion of the thin disk. Moreover, the study of low-redshift disk galaxies has revealed the presence of old stars dominating their stellar population in the outer disk regions \citep{yoachim2012,zheng2015}. Individual age determinations are in this context a key ingredient to test the viability of this scenario.

In the metal-poor thin disk, some of the slow rotator stars at $\textmd{R}_{\textmd GC}\geq 8$ kpc have metallicities of $\textmd{[Fe/H]}<-0.65$ dex. They are in a low number density region of the $\alpha$-metallicity space, which is partially an intersection of the metal-poor thick disk and halo sequences. In the same region, \citet{navarro2011} identified a group of stars with cold kinematics, interpreted as tidal debris from a disrupted dwarf galaxy. This is in the same line of the low$-\alpha$ accreted halo sequence described in \citet{nissen2010}. Because the three GMM groups partially intersect in this region and as a result of the errors in [Mg/Fe] and [Fe/H], we cannot reject the possibility that some of the stars at $\textmd{[Fe/H]}<-0.65$  in our metal-poor thin disk subsample correspond to the metal-rich tip of  the accreted halo misclassified as metal-poor thin disk. Alternatively, these stars could sample a true kinematically cold accreted component of the external thin disk. Even if this is the case, their small number is not enough to distort the results we found for the metal-poor thin disk.

Given the results of our analysis, it might be suggested that the metal-poor thin disk stars sample a special population of outer disk stars whose evolution might be different from the one on the inner disk. As shown by \citet{haywood2013} (their Fig. 17), these stars draw in the [$\alpha$/Fe] vs. Age plane a sequence parallel to the one of the local thin disk. As seen in Fig. \ref{fig:r_vs_vphi}, they rotate faster than the thick and thin disks in the solar neighborhood, decreasing toward thin disk values at larger radial distances. In addition, from Fig. 8 of \citet{guiglion2015}, it is possible to read a positive correlation of $\sigma _r$, $\sigma _\phi$ and $\sigma _Z$ with [Mg/Fe] for these stars. This evidence, if we assume that [Mg/Fe] is a proxy for the stellar age (but it is less strong than for the thick disk), can be seen as an indication of a population that was kinematically hotter in the past and cooled down during posterior evolution. In the same vein as the hypothesis considered by \citet{haywood2013}, this population could be composed of stars formed preferentially in the external regions of the thin disk, from a relatively low surface density gas distribution. If the gas was additionally polluted by the nucleosynthesis products of the chemical evolution of the thick disk, stars form at low metallicity from a pre-enriched medium of [$\alpha/$Fe] level comparable to the one of the thick disk. The relatively mild increase of velocity dispersion with [Mg/Fe] is compatible with a population evolving by self-enrichment at the same time as settling in a flatter structure while increasing its rotational speed. The indication of a  decrease of $|\textmd{Z}|$ with [Mg/Fe] visible in Fig. \ref{fig:distros_z} might be interpreted as a signature of an evolution in isolation, with few or no accretion events. This would agree with the dynamical analysis of \citet{ruchti2015}, who found scarce evidence for an accreted component in this region of the abundance-metallicity plane.

In principle, our results would also be compatible with the general chemical evolution idea of inside-out formation. In the outer regions, the longer infall timescales lead to less chemical enrichment
as a result of the lower gas density. The outer regions settle on a longer timescale, and consequently an increase of velocity dispersion and $|\textmd{Z}|$ with  age  would also be expected (as in any model that assumes the formation of the disk by slow gas accretion). As discussed in \citet{chiappini2001}, in the case of a double disk component (thick and thin understood as dual different star formation rates), the outer regions would  be contaminated by a previous component (thick disk/inner halo) because there the chemical enrichment has never been efficient enough. If the formation proceeds inside-out, then younger populations form with increasing scale length. This would imply a negative age gradient with Galactocentric distance, with  younger populations dominating at the outskirts of the disk.

Considering the limitations imposed by the scarcity of stronger observational constraints, in particular, of age distributions, it is difficult to distinguish between these two broad scenarios. If the external thin disk started to form stars by the end of the thick disk formation from material polluted by its nucleosynthesis products, its evolution could proceed almost dissociated from the inside-out formation taking place in inner regions. We could expect them to be composed mostly of old stars. In contrast, younger average ages are expected if the external disk is the product of a pure inside-out formation from a media contaminated by the halo or thick disk. Current evidence concerning  ages in the Milky Way \citep{haywood2013} and in external galaxies \citep{yoachim2012,zheng2015} seems to favor a formation proceeding in parallel to the inner parts of the thin disk.

All in all, our analysis leads us to characterize the metal-poor group of the thin disk as a distinct population, with different spatial and kinematic distributions with respect to those of the thick and the more metal-rich part of the thin disk. The size of our sample, and the intrinsic limitations of our data (difficulty of determining accurate distance, kinematics, and ages) prevent us from providing stronger detailed conclusions.

We hope that the new perspectives provided by the recent large spectroscopic surveys will trigger new theoretical efforts aiming at reconciling the observations with a new generation of chemodynamical models. Precise distances and proper motions from the forthcoming Gaia satellite and corresponding fundamental parameters from spectroscopic follow-up campaigns will help us to clearly define the spatial and kinematic distributions for stars in a larger spatial volume. The analysis of their correlation with chemistry and ages will open very promising avenues to construct a definitive coherent picture of the structure, formation, and evolution of the Galactic components.

%==================================================================================================
%    Conclusions
%==================================================================================================
\section{Conclusions}
\label{sec:conclusiones}

We made use of the iDR2 of the Gaia-ESO survey to probe the Galactic disk system in $\sim130$ lines of sight at low and high latitudes. We selected a subsample of 1725 stars by rejecting stars with low S/N values and large dispersions in stellar parameters and in abundance ratios [Fe/H] and [Mg/Fe]. This subsample spans a spatial region depicted in Fig \ref{fig:r_vs_z_restricciones}
that is mainly located in $5<{\textmd R}_{\textmd GC}<10$ kpc, and $\left|Z\right|<2$ kpc.

We demonstrated the usefulness of a clustering approach to separate subsamples from their distribution in the abundance-metallicity plane. This mathematical procedure allowed us to identify and characterize subgroups in data in a probabilistic
way, defined by changes in slope and/or variations in the star number density.

In summary, our main results and conclusions are as follows.

\begin{itemize}
 \item A Gaussian mixture model decomposition allowed us to find subgroups in the [Mg/Fe] vs. [Fe/H] plane (Fig. \ref{fig:gmm_vs_split_manual}). Five data groups were identified and associated with Galactic populations; the metal-rich end of the halo, the thick disk sequence, and three subgroups associated with the thin disk sequence. The separation between thin and thick disks is compatible with previous procedures based on estimating the minimum of the [Mg/Fe] distribution in metallicity bins.
 \item A comparative GMM analysis between our sample and a subsample of $\sim6100$ APOGEE-RC stars yielded qualitatively comparable results; the sequences and features observed in the abundance-metallicity plane, and as a consequence, the outcome of our GMM analysis in GES, might not be strongly biased by the selection function of the survey.
 \item The GMM decomposition does not depend on stellar types; qualitatively similar distributions were obtained for dwarf and RC subsamples. However, a different slope was found in the metal-rich end of the thin disk sequence. This could be explained as a consequence of the different radial regions sampled by the two groups of stars.
 \item The metal-poor-end group of the thin disk, with $\textmd{[Fe/H]}<-0.25$ dex, displays distinct properties with respect to the rest of thin disk metal-rich stars; it seems to have a qualitatively higher scale height and a larger rotational velocity component V$_\phi$ (Figs. \ref{fig:distros_z} and \ref{fig:r_vs_vphi}).
 \item Our findings, together with some other recent analyses of the Gaia-ESO iDR2 sample (\citeauthor{guiglion2015} \citeyear{guiglion2015}; \citeauthor{ruchti2015} \citeyear{ruchti2015}) and age considerations \citep{yoachim2012,haywood2013,zheng2015}, seem to favor a scenario in which the metal-poor thin disk formed in the outskirts, independently of or in parallel to the inside-out formation taking place in the inner regions of the disk. It might correspond to a self-enriched population, formed from pristine gas polluted by material expelled from the thick disk.  Posterior evolution led stars to settle from a relatively hot structure into a more disk-like one, increasing the azimuthal velocity and decreasing the velocity dispersions and scale height. The smooth changes of $|\textmd{Z}|$ and  V$_\phi$ with respect to [Mg/Fe] suggested by the data indicate a population that evolved in isolation, with few or no accretion events.
\end{itemize}

The chemical characterization of stellar populations has proven to be a fruitful way to separate stellar samples in a way that allows studying their unbiased kinematical and structural distributions. This provides physically meaningful insights to understanding their possible different chemodynamical histories.

In essence, our main results confirm the division of $\alpha-$ rich/poor sequences to define thin and thick disk populations devised in previous studies. At the same time, our further separation of a metal-poor group of thin disk stars allows us to characterize them as a special substructure. Its particular structural and kinematic properties reveal a population with a chemodynamical evolution possibly disconnected from that of the metal-rich thin disk.

The Gaia ESO survey is an ongoing project. Future releases will contain larger samples, providing the opportunity to test with a higher statistical significance the patterns imprinted by the specific evolution of Galactic stellar components, a step forward in our understanding of galaxy formation and evolution.

%==================================================================================================
%    Agradecimientos
%==================================================================================================
\begin{acknowledgements}
This work was partly supported by the European Union FP7 programme through ERC grant number 320360 and by the Leverhulme Trust through grant RPG-2012-541. We acknowledge the support from INAF and Ministero dell' Istruzione, dell' Universit\`a' e della Ricerca (MIUR) in the form of the grant "Premiale VLT 2012". The results presented here benefit from discussions held during the Gaia-ESO workshops and conferences supported by the ESF (European Science Foundation) through the GREAT Research Network Programme. A. Recio-Blanco, P. de Laverny and V. Hill acknowledge the Programme National de Cosmologie et Galaxies (PNCG) of CNRS/INSU, France, for financial support.
\end{acknowledgements}

%#########################################################################################################
%     BIBLIOGRAFIA
%#########################################################################################################

\bibliographystyle{aa}
\bibliography{biblio} 

\end{document}